\documentclass[pra,showpacs,twocolumn,showkeys]{revtex4-1}
\usepackage{graphics}
\usepackage{amsmath}
\usepackage{slashed}
\usepackage{mathtools}
\usepackage{color}
\usepackage{bbold}
\usepackage[usenames,dvipsnames]{xcolor}
\usepackage{dsfont}
\usepackage{braket}
\begin{document}
\newcommand{\bstfile}{osa}
\newcommand{\bibs}{d:/Dropbox/Dad/Mark/References/BibFile}
\title{Theoretical and experimental studies of photomechanical materials}
\author{Bojun Zhou, Elizabeth Bernhardt,  Ankita Bhuyan, Zoya Ghorbanishiadeh, Nathan Rasmussen, Joseph Lanska and Mark G. Kuzyk}
\email{kuz@wsu.edu}
\affiliation{Department of Physics and Astronomy, Washington State University, Pullman, Washington  99164-2814
\\ \today}
\begin{abstract}
After a brief introduction to the field of light-responsive materials, this paper provides a general theory for modeling the photomechanical response of a material, applies it to the two best-known mechanisms of photothermal heating and photo-isomerization, and then describes an experimental apparatus for quantitative measurements of the stress response.  Several different materials are characterized to illustrate how the experiments and theory can be used to isolate the contributing mechanisms both through photomechanical measurements and auxiliary measurements of laser heating and thermal expansion.  The efficiency and figure of merit of the photomechanical response is defined on several scales form the molecule to the bulk, and the photomorphon -- the basic material element that determines the bulk response -- is introduced.  The photomorphon provides a conceptual model that can be expressed in terms of viscoelastic elements such as springs in series and parallel with the photoactive molecule.  The photomechanical response, figure of merit, and the deduced microscopic photomechanical properties are tabulated and proposals for new materials classes are made.
\end{abstract}

\maketitle

\tableofcontents

\section{Introduction}\label{sec:intro}

Research on photomechanical materials, which change shape in response to light, has lead to many novel materials, a rich diversity of phenomena and fledgling applications.\cite{white17.01}  The first photomechanical effect was observed in the 1800s by Bell.\cite{ja:bell00.01}  Modern organic photomechanics was born in the early 1970s when Agolini and Gay serendipitously discovered that polymer doped with an azo dye is stressed by light.\cite{agoli01.01}  The first photomechanical device that combines the four device classes -- logic, transmission, sensing and actuation -- was reported in the early 1990s by Welker using an azo-dye-doped poly(methyl methacrylate) (PMMA) fiber\cite{welke94.01}.  Future-esque devices have been demonstrated\cite{welke96.02} as well as objects that morph into a variety of shapes, such as a flat sheet that folds into a box when exposed to light.\cite{Liue1602417}

Macroscopic light-induced changes of a material in the way of surface relief gratings\cite{viswa99.01,bian99.01,he00.01} ushered in an explosive growth in research of polymer-based photomechanical materials, leading to models of the underlying mechanisms.\cite{saphi05.01,tosch09.01}   Experimental studies of the underlying mechanisms that drive the photomechanical response and theoretical models are of a qualitative nature because measurements are imprecise.  In addition, there is no proposed figure of merit to compare materials aside from observations of how dramatic the motion appears in a video.\cite{bian06.01,serak09.01,nakan10.01}

This paper derives a broad and general framework that should be applicable to a wide range of materials, proposes a protocol for precise measurements of the photomechanical response and other relevant properties that are crucial for characterizing materials, defines a figure of merit for comparing their efficiencies, and introduces a procedure for isolating the mechanisms that contribute in a given material.  The concepts described here are applied to understanding how photomechanical changes at the molecular level affect successive hierarchies of size up to the observed bulk response.  This understanding is used to define figures of merit, depending on application -- and perhaps more importantly -- provides an intuitive way to view these processes by calling upon an energy surface at the molecular level that easily translates into spring models whose properties are affected by light.

A key concept is the introduction of the photomorphon (PM), the basic material unit of a photomechanical response, which is made of an active molecule and the surrounding correlated material that comprises the photomechanical unit (PU), and the surrounding polymer associated with a particular PU that together makes the PM -- the photomechanical ``unit cell."  A broad class of material classes and photomechanical mechanisms is shown to be able to be modeled in this way.

The experimental protocol is demonstrated and applied to several materials, which are characterized for their photomechanical response.  Models and auxiliary measurements are applied to extract the relevant parameters to isolate the mechanisms and to determine the figures of merit, giving a unified approach from theory to experiments and analysis.

\section{Overview}\label{sec:overview}

The action of light on a material to change its shape is a growing area of research due to the promise of directly converting light to mechanical energy, thus increasing conversion efficiency.  This paper seeks to model the origins of the bulk response in the underlying photomechanical units in a general way that can be applied to many materials and mechanisms.  Perhaps more importantly, this ground-up analysis can be applied to define figures of merit that select for desirable underlying properties that yield a usable photomechanical response.  For example, some researchers focus on maximizing the magnitude of the length change of the material while others study the forces.  Which is right?

\subsection{Photo-Heating as an Example}

The most obvious photomechanical effect is photothermal heating, where the material expands or contracts due to changes in temperature when the absorbed light energy is converted to heat. This simple example illustrates how one can calculate the efficiency of a photomechanical process by comparing the work done by an expanding rod to the energy of the absorbed photon, as follows.  We consider a microscopic view of photothermal heating, a regime in which we know that the bulk relationships we are using break down for a small number of molecules.  However, the results are good order-of magnitude estimates and thus worthwhile to consider.

The photothermal process starts with the absorption of a photon.\cite{Marsh13.01}  We select the smallest reasonable chunk of material that is large enough to be treatable classically and limit the temperature change to a few degrees.  Then, the increase in temperature is given by $\Delta T = dU/\rho V c$, where $dU$ is the photon energy absorbed, $\rho$ the mass density, $c$ the specific heat and $V$ the volume.  The fractional length change that results is given by $\Delta \ell / \ell = f \Delta T$, where $f$ is the coefficient of thermal expansion.  To determine the work done, we calculate the reverse process, that is, the work done in order to bring the rod back to its length prior to heating using the fact that for small train, the stress $\sigma$ and strain$ \Delta \ell/ \ell$ are related through Young's modulus $E$ according to $\sigma = E \Delta \ell / \ell$.

\begin{table}
\centering
\begin{tabular}{c c c}
  \hline
  Parameter & Silica Glass& PMMA \\ \hline
  Young's mod & E = 7.2 $\times$ 10$^{10}$ N/m$^2$ &  E = 3.3 $\times$ 10$^9$ N/m$^2$ \\
  Specific heat & c = 750 J/kg$\cdot$K & c = 1700 J/kg$\cdot$K \\
  Density & $\rho$ = 2200 kg/m$^3$ & $\rho$ = 1200 kg/m$^3$ \\
  Thermal Expand & f = 10$^{-6}$ K$^{-1}$ & f = 8$\times$10$^{-5}$ K$^{-1}$ \\
  \hline
\end{tabular}
\caption{Relevant photomechanical properties}
\label{tab:properties}
\end{table}

Putting this all together, the work done is given by
\begin{align}\label{eq:HeatWork}
W = \frac {1} {2} \frac {E} {V} \left(\frac {f \, dU} {\rho c} \right)^2 .
\end{align}
Table \ref{tab:properties} gives ballpark values of the material parameters for silica glass and poly (methyl methacrylate) (PMMA) polymer.  We assume that the photon energy is $dU = 3.8 \times 10^{-19} J$ (wavelength of 522 nm) -- green in the visible spectrum, and that the material volume is $4 \times 10^{-26} m^3$, which keeps the temperature change to a few degrees.  Plugging the numbers into Equation \ref{eq:HeatWork} gives $W_\text{silica} = 5 \times 10^{-26} J$ for an energy conversion efficiency per absorbed photon of $\xi = W_\text{silica}/dU = 3.8 \times 10^{-7}$ and $W_\text{PMMA} = 9 \times 10^{-24} W$ for a conversion efficiency of $\xi = W_\text{PMMA}/dU = 2.8 \times 10^{-5}$.

One might imagine designing materials that have larger $f$, $dU$, and $E$ with smaller $c$ and $\rho$.  This avenue might prove fruitful in increasing the efficiency, but dissipating heat is often an issue and the heating response tends to be slow.  The models that we develop below apply to heating and other mechanical mechanisms.  The hope is that the heating efficiencies can be greatly surpassed with other mechanisms, but -- as we later show -- this has not yet become reality.

We round out this section by outlining the general ideas, which are further developed in the sections that follow using an intuitive spring model that applies to both single molecules and bulk materials.

\subsection{Mechanical Spring Model}\label{sec:MolecularSringModel}

\begin{figure}\centering
    \includegraphics{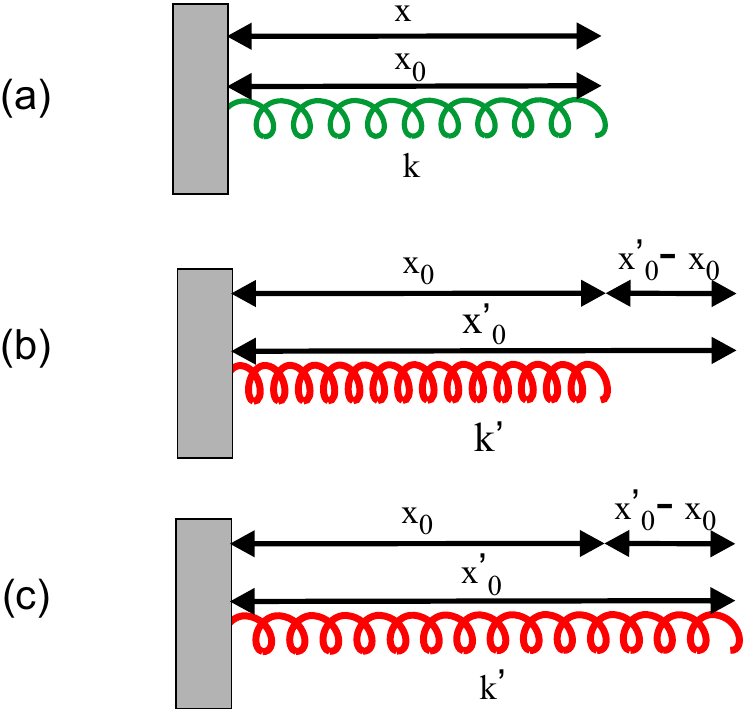}
    \caption{A spring model of the photomechanical response.  (a) The resting spring (b) immediately after it is excited by light and (c) after relaxing to its new length. }\label{fig:PMunit}
\end{figure}

Figure \ref{fig:PMunit} shows a spring model of the photomechanical response.  This unit is general and could be a chunk of material, a liquid crystal domain, a molecule, interconnected molecules/polymers, etc.  The only important qualities is that it have a viscoelastic response and that light affects these properties.  When stretched by a force $F$, its potential energy is given by
\begin{align}\label{eq:PotEnergy}
V(x) = \frac {1} {2} k \left(x - x_0 \right)^2 ,
\end{align}
where $x_0$ is the resting length of the spring and $k$ its force constant.  Note that the force constant is generally complex, where the imaginary part gives damping, but here we focus on the real part.

Now consider a photon that excites the spring.  Central to the mechanical spring model is a photon-induced change of the spring constant and its equilibrium length.  We will see in the next section how this picture naturally arises from the model of molecular vibrations.  If the spring starts from its resting state with $x = x_0$ so the parameters change to $x_0 \rightarrow x_0^\prime$ and $k \rightarrow k^\prime$, the energy increase will be given by
\begin{align}\label{eq:PotEnergyChangeRest}
dV = \frac {1} {2} k^\prime \left( x_0^\prime - x_0 \right)^2 \le \hbar \omega,
\end{align}
where -- for a reversible process -- energy conservation demands that the photon energy $\hbar \omega$ must exceed the energy imparted to the system.  Equation \ref{eq:PotEnergyChangeRest} implies that 100\% efficiency is possible in a system where the spring's properties are strategically chosen.  Note that the parameters $x_0$, $x_0^\prime$, $k$ and $k^\prime$ can be calculated for a molecule from quantum principles.

\begin{figure}\centering
    \includegraphics{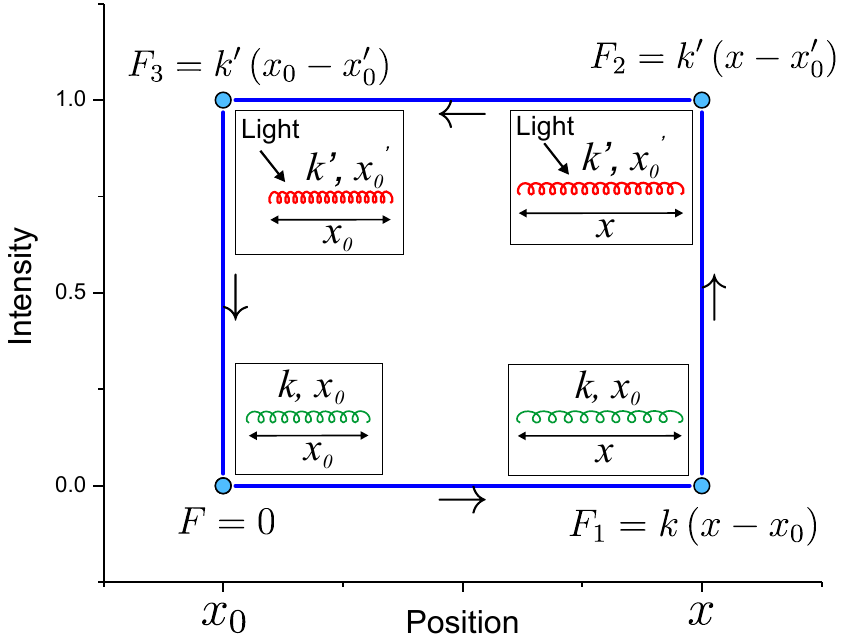}
    \caption{The spring starts at rest so $F=0$ and goes through a cycle where it is (1) stretched to $x$ with Force $F_1$; (2) the intensity is turned on requiring a force $F_2$ to keep the spring's length fixed as its spring constant and equilibrium changes, (3) a compressional force brings the spring back to its original length in the presence of the light.}\label{fig:IxCycle}
\end{figure}

Before proceeding, we show how an experiment can determine $k^\prime$ and $x_0^\prime$.  Figure \ref{fig:IxCycle} shows a cycle starting with no applied force ($F = 0$) in its equilibrium state of length $x_0$.  Then, the spring is stretched to a length $x$ by an applied force $F_1 = k\left( x - x_0 \right)$.  Since the applied force is controllable and the length change measured, the spring constant can be determined from the two known properties
\begin{align}\label{eq:k-determine}
k = \frac {F_1} {x - x_0}.
\end{align}
This is how Young's modulus is measured in a bulk chunk of material.

Then, when light illuminates the spring, causing the equilibrium length to change to $x_0^\prime$ and the spring constant to become $k^\prime$, the applied force required to keep the length fixed is
\begin{align}\label{eq:F2-Equation}
F_2 = k^\prime \left( x - x_0^\prime \right).
\end{align}
Since both $x_0^\prime$ and $k^\prime$ are unknowns, this equation alone cannot be used to determine them.  To do so, the spring is compressed to its original length while the beam is on, requiring a force
\begin{align}\label{eq:F3-Equation}
F_3 = k^\prime \left( x_0 - x_0^\prime \right).
\end{align}

Equations \ref{eq:F2-Equation} and \ref{eq:F3-Equation} can be used together to determine the remaining parameters, yielding
\begin{align}\label{eq:k-prime}
k^\prime = \frac {F_2 - F_3} {x - x_0}.
\end{align}
and
\begin{align}\label{eq:x0-prime}
x_0^\prime = \frac {F_2 x_0 - F_3 x} {F_2 - F_3}.
\end{align}
Equations \ref{eq:k-prime} and \ref{eq:x0-prime} can be substituted into Equation \ref{eq:PotEnergyChangeRest} to determine the energy gained when exciting the spring in terms of the measured parameters during the cycle given by Figure \ref{fig:IxCycle}, yielding
\begin{align}\label{eq:dV-Measured}
dV = \frac {1} {2} \frac {F_3^2} {F_2 - F_3} \left( x-x_0\right).
\end{align}
Determining this energy does not require  the properties of the initial spring because it was not stretched.  If it had been, $k$ would have been required, which is determined from Equation \ref{eq:k-determine}.

Equation \ref{eq:dV-Measured} gives the energy provided to the spring by the light in terms of the measured forces applied and displacements measured.  A more special case is the one where the spring constant does not change appreciably ($k \approx k^\prime$) and the photomechanical energy derives solely from a change in the resting length.  In this case, the change in energy upon exciting the molecule is given by
\begin{align}\label{eq:dV-k=k'}
dV = \frac {1} {2} k \left( x_0^\prime - x_0 \right)^2 = \frac {1} {2} \frac {1} {k} \left( F_1 - F_2 \right)^2 ,
\end{align}
where the second equality in Equation \ref{eq:dV-k=k'} is derived by expressing $x_0 - x_0^\prime $ in terms of the forces $F_1$ and $F_2$ and the spring constant $k$.  Note that when the spring constants are the same, only forces $F_1$ and $F_2$ need to be measured provided that $k$ is known from a measure of Young's modulus.

The cycle shown in Figure \ref{fig:IxCycle} is an indication that the system can be used as a motor that operates with constant illumination.  An example of such motors using photomechanical ``belts" wrapped around wheels to make an engine\cite{kneze13.01} and a turbine configuration\cite{kneze14.01} have been reported by Kne\v{z}evi\'{c} and Warner.

The spring model presented here forms the basis of all that follows.  The thermal model requires the light to remain on to keep the material in its excited state.  The mechanical spring model as presented assumes that the molecule will quickly de-excite, in which case another photon will need to be supplied to re-excite it.  However, molecules with long-lived excited states can persist in the excited mechanical state.  Then, the molecule can be forced to de-excite using a second photon that is tuned to the appropriate wavelength.  Thus, the molecule can be toggled between two states.  Collections of molecules, on the other hand, must follow population models, which behave analogously to the thermal model.

\section{Theory}\label{sec:theory}

This section builds a material from the ground up.  In essence, springs in series and parallel model active molecules and passive elastic elements such as a polymer host.  Section \ref{sec:ActiveMolecule} presents a classical spring model of an active molecule whose mechanical properties change under the action of the light.  Section \ref{sec:PMunit} introduces the photomechanical unit (PM unit) as a passive element in parallel with the active molecule, which partially impedes its motion as is typical of composite systems.  The PM unit can represent any multi-component system such as an active molecule in a passive polymer host.  Furthermore, PM units can be attached to each other through additional passive units, which can be modeled with an elastic element in series with the PM units.  Section \ref{sec:photomorphon} defines the photomorphon, which is the smallest part of a material that includes one active molecule, the passive restraint and the connective element.  A photomorphon is the smallest part of a material that behaves as the bulk in that mechanical state, and is the basis of statistical models of a material's photomechanical response.  The mechanical properties of all the material elements, from molecule to photomorphon can be associated with the mechanical spring model presented in Section \ref{sec:MolecularSringModel}.  Since the photomorphon is the basic building block whose efficiency defines the macroscopic efficiency, we will apply the loop shown in Figure \ref{fig:IxCycle} to the photomorphon.

The photomorphon can be in its resting or excited state.  Section \ref{sec:StatModel} shows how the bulk mechanical properties of the material are determined from a weighted ensemble average over the properties of the photomorphons in these two possible states.  The equilibrium populations, in turn, are modelled from statistical mechanical considerations.  Section \ref{sec:PopDynamics} describes how the dynamics of the populations are driven by photons that excite the photomorphon and relaxation processes that bring the photomorphon back to its lowest-energy resting state through both spontaneous and stimulated decay.

The photomorphon is the ``unit cell" of the material that is characterized by its resting length $x_0$, force constant $k$, excited length $x_0^\prime$ and excited force constant $k^\prime$.  Thus, only these four phenomenological parameters are needed without detailed information about the material's composition.  Alternatively, these parameters can be determined from a microscopic treatment of the material, making it possible to test the underlying mechanisms and theories of the response. Insights gained into how microscopic forces and structures can be used to control the bulk properties can be applied to designing the ideal photomechanical material.

Section \ref{sec:MicroToMacro} describes how the photomorphon's properties translate into the bulk response and Section \ref{sec:StressToStrain} shows how the clamped stress response is related through Young's modulus to the unclamped strain response.  Section \ref{sec:PMefficiency} proposes a simple definition of a figure of merit that describes the efficiency of a material in converting light energy to mechanical work and Section \ref{sec:DyeDopedPolyExample} ends the theory section with a mechanical model of a dye-doped polymer.

\subsection{The Active Molecule}\label{sec:ActiveMolecule}

\begin{figure}\centering
    \includegraphics{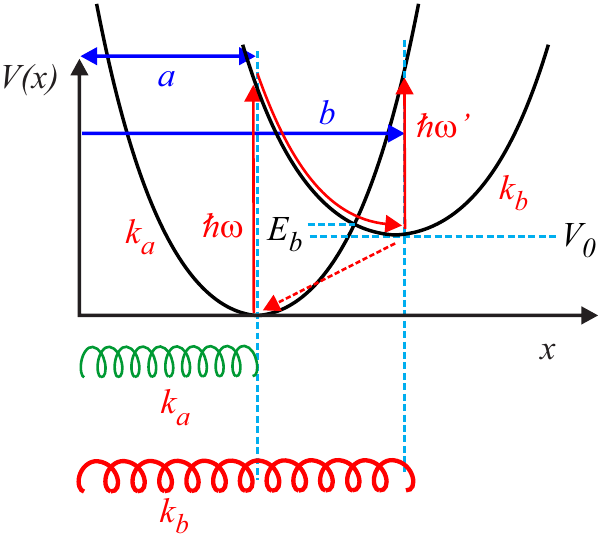}
    \caption{A molecule can be modeled as a spring of force constant $k_a$ and resting length $a$ which has a spring constant of $k_b$ and resting length $b$ upon excitation with light.}\label{fig:SpringsModel}
\end{figure}

Consider an energy-level diagram of a photomechanical spring, which represents a classical view of a molecule, as shown in Figure \ref{fig:SpringsModel}.  The ground state spring constant $k_a$ and equilibrium positions of the nuclei quantified by $a$ are governed by chemical bonds, which originate in the electron cloud shared by the nuclei.  Upon excitation by light of energy $\hbar \omega$, the electrons in the molecules rearrange in a way that changes both the spring constant and the resting length of the spring, as represented by the upper parabola centered at $x=b$ and of curvature $k_b$.  Since the massive nuclei respond slowly, the length of the molecule just after it is excited is initially of length $x=a$.  Subsequently, the length relaxes to $b$.  It is this relaxation process that expands the spring from $x=a$ to $x=b$, doing work on its surroundings.

Classically, the length of the spring would remain indefinitely at $x=b$ because the barrier $E_b$ must be overcome to bring the system back to the ground-state parabola.  Quantum mechanically or thermally, the spring can ``tunnel" through this barrier  and then relax into its de-excited state, leaving the system in its ground state.  The random nature of tunneling precludes doing useful work.  The quantum or thermal nature of tunnelling is irrelevant to the model we develop below, which will treat the tunnelling rate as a phenomenological parameter.  Typically, the excitation rate greatly exceeds the tunneling rate, so that a large population of excited state molecules is generated.  As we will show, the population ratio for a given excitation intensity can be determined from the ratio of the excitation and relaxation rates.

In one full cycle, energy $\hbar \omega$ is provided to the system and useful work $\hbar \omega - V_0$ is performed.  Clearly, minimizing $V_0$ maximizes the efficiency.  However, the parameters in this spring model are related to each other because the work done by the excited molecule in transitioning from length $a$ to $b$ is governed by the energy conservation condition
\begin{align}\label{eq:WorkDone}
\hbar \omega - V_0 = \frac {1} {2} k_b (a-b)^2.
\end{align}

It is possible to force the molecule to relax to its resting state from the excited state using a photon.  As shown in Figure \ref{fig:SpringsModel}, a photon of energy $\hbar \omega^\prime$ can excite the molecule to the energy surface of curvature $k$, thus allowing the molecule to relax back to its resting state.  using such a ``stimulating" photon will accelerate the decay to the resting state.  In addition, the molecule can do work on its environment as it relaxes, thus increasing the speed of the process and doing more work.

The work done by the first photon is $\hbar \omega - V_0$ while the work done by the second photon is $\hbar \omega^\prime + V_0$, yielding total work
\begin{align}\label{eq:TPAWorkDone}
dV = \left( \hbar \omega - V_0 \right) + \left( \hbar \omega^\prime + V_0 \right) = \hbar \omega + \hbar \omega^\prime.
\end{align}
As such, the efficiency of this process is 100\%, where the full energy of the light is converted into work.  However, it would be difficult to harness such work because -- in the first part of the cycle -- the molecule is expanding while -- in the second half -- it is contracting.  In such a closed loop, no net work is done by a conservative force.  But, a clever configuration might be devised that takes advantage of such a cycle.

Another limiting factor on the efficiency is the small fraction of absorbed photons that lead to a length change.   The efficiency of a process requiring the absorption of two separate photons is proportional to the joint probability, thus further lowering the efficiency.  These factors must be taken into account when designing an efficient photomechanical device.

\subsection{The Photomechanical Unit and its Efficiency}\label{sec:PMunit}

The efficiency of the photomechanical response, defined as the useful work done on the external world, will depend on the efficiency of the PM unit as well as the properties of the photomorphon (described later).  Here we focus on the PM unit.

\begin{figure}\centering
    \includegraphics{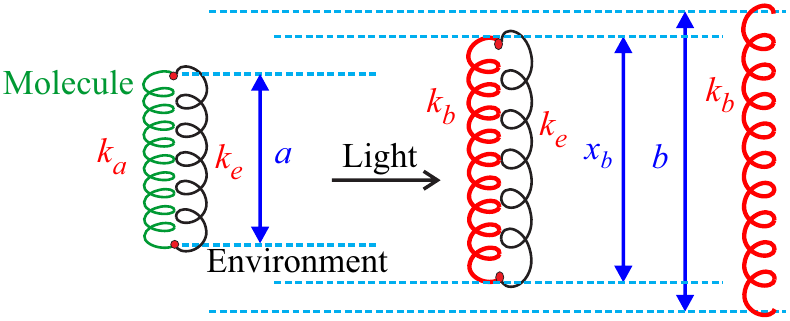}
    \caption{A photoactive molecule in parallel with a passive element equilibrates to length $x_b$.  The passive element prevents the dye from fully expanding upon excitation to its excited length $b$.}\label{fig:PMunitEff}
\end{figure}

Consider the simplest example of a PM unit made with an active molecule of spring constant $k_a$ and a passive part due to the environment in parallel with it having spring constant $k_e$, both in their resting state of length $a$ as shown in Figure \ref{fig:PMunitEff}.  Upon excitation, the active molecule's length changes to length $b$ and the spring constant becomes $k_b$.  However, the passive environment will restric the length change, leading to an equilibrium length $x_b$, where $a< x_b < b$, which is calculated from a balance between the two spring forces, or
\begin{align}\label{eq:BalanceSprings}
k_b \left(b - x_b \right) = k_e \left(x_b -a \right),
\end{align}
where the passive spring with force constant $k_e$ is stretched by $x_b - a$ from its equilibrium value and the active spring of force constant $k_b$ by an amount $b - x_b$.

Solving Equation \ref{eq:BalanceSprings} for $x_b$ yields
\begin{align}\label{eq:PMunitEquil}
x_b = \frac {b k_b + a k_e} {k_e+k_b} .
\end{align}
Using Equation \ref{eq:PMunitEquil}, the resting length of the parallel spring combination is
\begin{align}\label{eq:PMunitEquil(x0-a)}
x_b - a = \frac {(b-a)k_b} {k_e + k_b} .
\end{align}
Also
\begin{align}\label{eq:PMunitEquil(b-x0)}
b - x_b = \frac {(b-a)k_e} {k_e + k_b} .
\end{align}
Using Equation \ref{eq:PMunitEquil(x0-a)}, the potential energy of the two-spring system upon excitation of the photomechanical molecule is given by
\begin{align}\label{eq:PMunitEEnergy}
V(k_e) = \frac {1} {2} \left( k_b + k_e \right) \left( x_b - a \right)^2 = \frac {1} {2} \frac {k_b^2} {k_e + k_b} \left(b - a \right)^2 ,
\end{align}
where we have used the fact that the spring constant of the PM unit in the excited state is given by
\begin{align}\label{PMunitGroundK}
k_b^\text{eff} = k_b + k_e.
\end{align}
In the ground state, the PM unit's spring constant is
\begin{align}\label{PMunitGroundK}
k_a^\text{eff} = k_a + k_e.
\end{align}

Equation \ref{eq:PMunitEEnergy} is the energy available to do work.  In the absence of the spring $k_e$, the full energy of the photomechanical potential $V_0 \equiv V(0)$ from the molecule is available to do work, so we can define the PM unit efficiency by
\begin{align}\label{eq:PMunitEffic}
\xi_\text{PM} = \frac {V(k_e)} {V_0} = \frac {k_b} {k_e + k_b} .
\end{align}
As we would expect, the efficiency of the PM unit given by Equation \ref{eq:PMunitEffic} is unity when $k_e = 0$ and gets smaller as $k_e$ increases and more useful energy is lost to the parasitic environment.  This analysis thus shows that the best photomechanical material is one made of a dense network of interconnected photoactive molecules without a passive environment.

We stress that this is yet another source of loss, in addition to the intrinsic losses in the active molecule as described at the end of Section \ref{sec:ActiveMolecule}.

\subsection{The Photomorphon}\label{sec:photomorphon}

\begin{figure}\centering
    \includegraphics{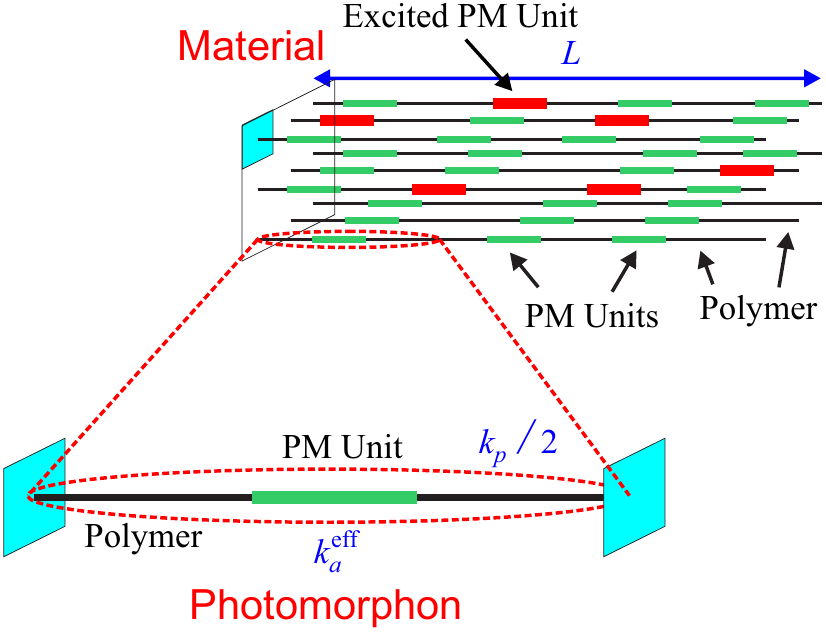}
    \caption{A bulk material made of polymer chains and photomechanical units.  The smallest photomecahnical part that retains the properties of the bulk material is called a photomorphon.}\label{fig:Photomorphon}
\end{figure}

\begin{figure}\centering
    \includegraphics{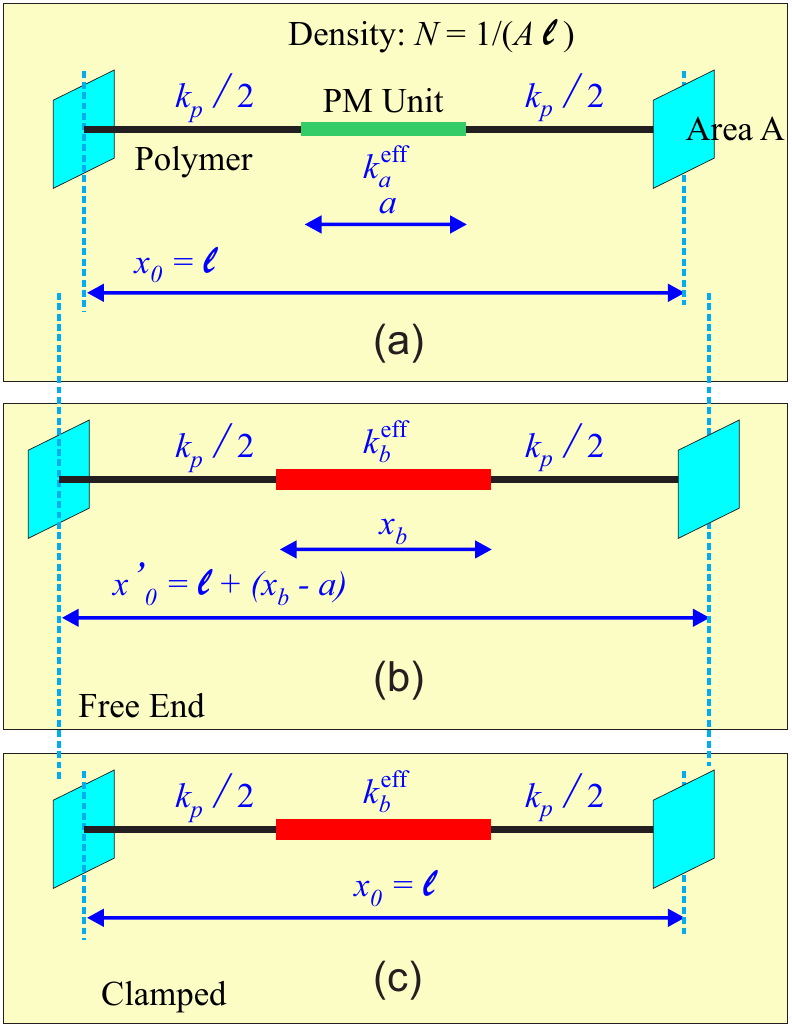}
    \caption{(a) The ``average" unit consists of one photomechanical unit of length $a$ with spring constant $k_a^\text{eff}$ on a chain of length $\ell$ with total polymer spring constant $k_p$. (b) The excited state photomorphon with free ends and (c) the clamped photomorphon.}\label{fig:Configurations}
\end{figure}

When the photoactive material is dispersed into a host material, such as a dye into a polymer matrix, the section of the molecule that interacts with the polymer will be the PM unit.  For simplicity, we picture the polymer as one-dimensional chains as shown in Figure \ref{fig:Photomorphon}.  Figure \ref{fig:Configurations}a shows the average unit with one PM unit attached to a segment of length $\ell$.  We call this average unit the photomorphon, which is the unit cell of a photomechanical material.  It contains a PM unit and host material in which it is embedded.  The concept of a photomorphon applies to a broad variety of materials.

\begin{figure}\centering
    \includegraphics{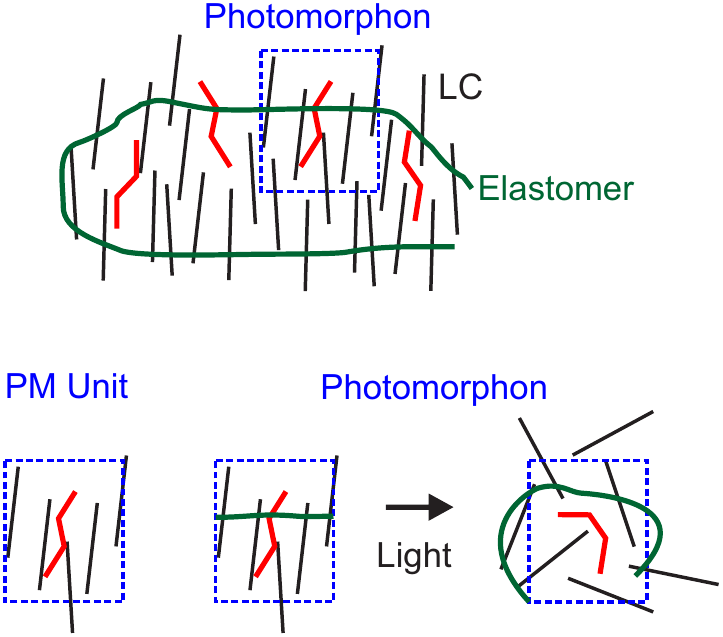}
    \caption{A dye-doped liquid crystal elastomer is made of liquid crystal units that are attached to an elastomer chain. Molecules that change conformation are added to the material and align themselves with the liquid crystal.  A dye molecule with surrounding mesogens make the PM unit and the PM unit with the attached elastomer is the photomorphon.  Light changes the conformation of the molecule from trans to cis, interfering with the liquid crystalline ordering and deforming the elastomer, thus changing the material's properties. }\label{fig:LC-Elastomer}
\end{figure}

One can imagine a photomorphon being made of photoactive molecules embedded in a glassy polymer, as one has for a dye-doped polymer or a polymer with the photoactive molecules covalently attached.  Even a more complex system such as a dye-doped liquid crystal elastomer, as shown in Figure \ref{fig:LC-Elastomer}, can be modelled as a photomorphon.  The active molecule used in many photomechanical materials are the class of azo-benzene dyes,\cite{priim12.01,priim14.01} which change from a trans to cis isomer upon excitation.

The PM unit is then one dye molecule surrounded by a liquid crystal in its nematic phase, and the liquid crystal acts as the hindering environment.  If the photon changes the shape of the dye, in this case from the trans to the cis conformation, it will interfere with the aligning forces, causing a decrease in the orientational order of the liquid crystal, and thus also affecting the mechanical properties of the PM unit.  The equilibrium shape of the PM unit will change, as does its elasticity.  The photomorphon, which includes the elastomer chain, will then deform and reflect the bulk photomechanical response.

In the one-dimensional chain model, $n = 1/\ell A$ is the number density of dye, which occupies the fraction $a/\ell$ of the photomorphon's length and therefore occupies that same fraction of the material's length.  The photomorphon has the same number density as the dye.  Since the material is prepared in this state, both the molecule and the polymer are in their resting length state.

The average PM unit shown in Figure \ref{fig:Configurations}a has an effective resting spring constant of $k_a^\text{eff}$ and resting length $a$, which are determined form the composition and structure of the material.  For example, the dye molecule might be in parallel with part of the polymer or might straddle two chains.

The spring constant of the polymer on both sides of the PM unit is $k_p/2$, which -- when taken in series -- gives a spring constant of $k_p$.  Then, the spring constant $k$ of the resting photomorphon is given by
\begin{align}\label{eq:GroundSpring}
k = \frac {k_p k_a^\text{eff}} {k_p + k_a^\text{eff}} .
\end{align}
Similarly, the excited state spring constant of the photomorphon is given by
\begin{align}\label{eq:ExcitedSpring}
k^\prime = \frac {k_p k_b^\text{eff}} {k_p + k_b^\text{eff}} .
\end{align}

We are now interested in calculating the photomechanical response of the the photomorphon upon excitation of the PM unit.  The ends of the photomorphon can be free to move as in Figure \ref{fig:Configurations}b or the ends can be clamped, as shown in Figure \ref{fig:Configurations}c.  Thus the clamped configuration yields uniform photomechanical stress with constant length while the unclamped length change of the photomorphon equals the length change of the PM.

\subsection{Statistical Model}\label{sec:StatModel}

In the steady state, a material will be made of a collection of resting and excited photomorphons. This section determines the bulk material properties from a weighted population average to determine various quantities such as Young's modulus and the photomechanical constants.  Our approach is to first determine the properties of the ``average" photomorphon, which is easily related to the bulk properties.  Then, population dynamics models can be used to determine how these bulk properties change with light exposure or temperature change.

\subsubsection{Force}

\begin{figure}\centering
    \includegraphics{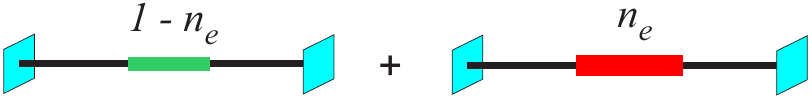}
    \caption{The ``average" photomorphon's properties will be the population-weighted average. }\label{fig:AveragePhotomorphon}
\end{figure}

If the excited state population fraction is given by $n_e$, then the ground state population fraction is  $1-n_e$, so the average photomorphon length in the unclamped configuration, $\bar{\ell}$, is given by
\begin{align}\label{eq:AverageChainSlength}
\bar{\ell}(\ell) = \ell (1 - n_e) + \left(\ell + \left(x_b - a \right) \right) n_e = \ell + n_e (x_b - a).
\end{align}

Similarly, the population-weighted average force constant for photomorphons connected in series is
\begin{align}\label{eq:AverageChainSpringConstant}
\frac {1} {K(n_e)} = \frac {n_e} {k^\prime} + \frac {1 - n_e} {k} \rightarrow K(n_e) = \frac { k} {1 - n_e \left( 1 - \frac {k} {k^\prime} \right) } .
\end{align}

The change in the resting length upon excitation from the ground state is
\begin{align}\label{eq:PopRestLength}
\delta \ell (n_e) = n_e \left( x_b - a \right)
\end{align}
giving a strain
\begin{align}\label{eq:PopRestAveStrain}
u(n_e) = \frac {\delta \ell} {\ell} = n_e \frac {\left( x_b - a \right)} {\ell}
\end{align}
and the force on the walls in the clamped configuration will be given by
\begin{align}\label{eq:PopAveForce}
F(n_e) = K(n_e) \delta \ell =  \frac {n_e  k \left( x_b - a \right)} {1 - n_e \left( 1 - \frac {k} {k^\prime} \right) } ,
\end{align}
where we have used Equations \ref{eq:AverageChainSpringConstant} and \ref{eq:PopRestLength}.  Other quantities of interest can be similarly calculated.

\subsubsection{Population Dynamics}\label{sec:PopDynamics}

In a real system, the temperature will set the equilibrium population of excited and ground state molecules in the absence of light.  The light will then drive excitations that change the populations, which decay back to equilibrium once the light is turned off.  Section \ref{sec:StatModel} gives the force and length change for an ``average" photomorphon.  Here we will calculate the time evolution of the populations, which can be applied to determining the time dependence of the forces and displacements.

Consider a region of material with $N$ photomorphons where $N_e$ of them are in their excited states.  Briefly, we consider populations $N$ instead of number densities $n$ because it simplifies the concepts.   Light will convert the lower energy state to the higher energy one; but also, the reverse process of light-stimulated de-excitation is possible.  In addition, the excited state population will decay in the absence of light to its equilibrium population.  Putting it all together, the time evolution of the excited state population is
\begin{align}\label{eq:PopTimeEvolve}
\frac {dN_e} {dt} = \left( N- N_e \right) \left( \frac {\eta_0 P} {\hbar \omega}  \right) - N_e \left( \frac {\eta_e P} {\hbar \omega}  \right) - \beta N_e ,
\end{align}
where $\eta_0$ is the probability that an incident photon converts a ground-state photomorphon to its excited state, $\eta_e$ the probability of converting an excited state photomorphon to its ground state, $P$ the {\em constant} power of the light that is turned on at $t=0$ and $\hbar \omega$ the photon energy.  Thus $P/\hbar \omega$ is the number of photons per unit time incident upon the material and $\beta$ is the population decay rate in the absence of light.

Defining the cross section $\sigma_i =  A \eta_i $ for state $i$, where $A$ is the area of the beam, and re-expressing Equation \ref{eq:PopTimeEvolve} in terms of the intensity $I = P/A$ after some re-arrangement of the terms yields
\begin{align}\label{eq:PopTimeEvolve2}
\frac {dN_e} {dt} = N I \left( \frac {\sigma_0} {\hbar \omega}  \right) - N_e  I \left( \frac {\sigma_0 + \sigma_e} {\hbar \omega}  \right) - \beta N_e .
\end{align}
Equation \ref{eq:PopTimeEvolve2} is of the form
\begin{align}\label{eq:PopTimeEvolveSimple}
\frac {dn_e} {dt} = \alpha _0  I -  n_e \left( \alpha I + \beta \right)
\end{align}
where we have divided Equation \ref{eq:PopTimeEvolve2} by $N$ to get the fractional number density $n_e = N_e/N$, and
\begin{align}\label{eq:ParametersDef}
\alpha_0 =  \frac {\sigma_0} {\hbar \omega} \hspace{2em} \mbox{ and } \hspace{2em} \alpha = \frac {\sigma_0 + \sigma_e} {\hbar \omega} .
\end{align}
Note that in the absence of photon-induced de-excitation, $\alpha = \alpha_0$.

Integration of Equation \ref{eq:PopTimeEvolveSimple} gives
\begin{align}\label{eq:PopTimeEvolveSimpleInegrate}
\frac {\ln \left(  \alpha _0 I -  n_e \left( \alpha I + \beta \right) \right)} {- \beta - \alpha I} = t + C ,
\end{align}
where we determine the integration constant $C$ under the condition that $n_e(0) = n_e^{(0)}$, yielding
\begin{align}\label{eq:PopTimeEvolveSimpleIntConst}
C = \frac {\ln \left(  \alpha _0 I -  n_e^{(0)} \left( \alpha I - \beta \right) \right)} {- \beta - \alpha I} .
\end{align}
We note that $n_e^{(0)}$ is the equilibrium number density of excited photomorphons in the dark, where the temperature dependence can be calculated from the partition function, yielding \begin{align}\label{eq:PopEquil}
n_e^{(0)} (T) = \frac {1} {1 + e^{V_0/k_B T}} ,
\end{align}
where $V_0$ is the energy difference between the cis and trans isomer.

Putting it all together, we get the time dependence of the excited state population of photomorphons after the light is turned on
\begin{align}\label{eq:PopTimeTurnOn}
n_e (t) = \frac { \alpha_0 I - \left(  \alpha _0 I -  n_e^{(0)} \left( \beta + \alpha I \right)  \right) e^{- (\beta + \alpha I)t} } {\beta + \alpha I}  .
\end{align}
After the light is turned off, the excited state population will decay according to
\begin{align}\label{eq:PopTimeEvolveTurnOff2}
n_e (t) = n_e^{(0)} (T) + (n_e (t_0) - n_e^{(0)} (T)) e^{- \beta (t - t_0) }  ,
\end{align}
where $n_e(t_0 )$ is the population of excited state photomorphons at the instant the light is turned off after it had acted for a time $t_0$.  Note that the rise-time constant with the light on depends on the power while the decay time constant depends only on $\beta$.  At low enough temperatures where the cis population is small,
\begin{align}\label{eq:PopTimeEvolveTurnOff}
n_e (t) =  n_e (t_0) e^{- \beta (t - t_0) }  .
\end{align}

Liquid crystal elastomers have been shown to have a time constant independent of power.\cite{harve07.01}  This implies that $\beta \gg \alpha I$ for the power range used in those experiments.  If this is so, Equation \ref{eq:PopTimeTurnOn} yields the long-time amplitude
\begin{align}\label{eq:PopTimeInfinity}
n_e (t\rightarrow \infty) = \frac { \alpha_0 I } {\beta + \alpha I} \approx \frac {\alpha_0 I} {\beta} - \frac {\alpha_0 \alpha I^2} {\beta^2},
\end{align}
where Equation \ref{eq:PopTimeInfinity} yields $n_e (t\rightarrow \infty) = I \alpha_0 /\beta$ when $\beta \gg \alpha I$.  In this limit, the amplitude is a linear function of intensity with slope $\alpha_0 /\beta$.  Thus a measure of the time constant, which is approximately given by
\begin{align}\label{eq:PopTimeConstInfinity}
\tau = \frac {1} {\beta + \alpha I} \approx \frac {1} {\beta} - \frac {\alpha} {\beta^2} I
\end{align}
gives $\beta$ and the linear part of the long-time photomechanical response  gives $\alpha_0 /\beta$, so these two values can be used to determine $\alpha_0$.  The parameter $\alpha$ can be determined at higher powers where the amplitude and time constants depend on power.  In this high-power limit,
\begin{align}\label{eq:PopTimeInfinityHighPow}
\tau = \frac {1} {\alpha I}
\end{align}
and
\begin{align}\label{eq:PopnfinityHighPow}
n_e (t \rightarrow \infty) = \frac {\alpha_0} {\alpha}.
\end{align}
Section \ref{sec:MicroToMacro} describes how macroscopic measurements of the length change are used to get this information.

\subsection{Pure Heating}

Pure heating is the process in which the photon's energy is deposited into the material and the temperature increase drives thermal expansion.  These effects  can be small due to typical thermal expansion coefficients of $\delta L/L \approx 10^{-5} $ to $10^{-4}$ per degree Kelvin temperature increase. The thermal expansion of a liquid crystal elastomer is much more dramatic, with $\delta L/L \approx 4 \times 10^{-3}$ per degree kelvin at room temperature, and increasing by orders of magnitude when the temperature nears the the transition from the nematic to isotropic phase.\cite{tajba01.01}  As such, photo-heating might make a large contribution the photomechanical response.  This section determines the dynamics of such a process.

We start by calculating the temperature change in response to energy deposited by light to a thin sample whose planar area $A$ is uniformly illuminated.  The heat equation using light as the source is
\begin{align}\label{eq:HeatEq}
\frac {\partial T} {\partial t} & = - \gamma(T-T_0) + C_p^{-1}(T)A A^\prime(\lambda) I \nonumber \\
& \equiv - \gamma(T-T_0) + C_p^{-1}(T) P_\lambda,
\end{align}
where $T$ is the sample temperature, $T_0$ is the ambient temperature, and $\gamma$ the rate at which the sample cools to ambient.  $\gamma$ will depend on the material as well as the sample's shape and size.  The intensity of light $I$ deposits energy at a rate $P_\lambda \equiv A A^\prime (\lambda) I$, where the absorption coefficient $A^\prime(\lambda)$ depends on the wavelength of light $\lambda$.  $C_p (T)$ is the heat capacity, which depends on the temperature.

We expand the heat capacity in a series of the temperature difference $T-T_0$ to first order in temperature
\begin{align}\label{eq:ExpandCapacity}
\frac {1} {C_P}  = c_0 - c_1 \left(T-T_0 \right),
\end{align}
where $c_0 = C_p^{-1}(T_0)$.  Note that we have added the negative sign because -- with this definition -- $c_1>0$ when the heat capacity increases with temperature.  At long times, when $\partial T / \partial t = 0$, Equation \ref{eq:HeatEq} with the help of Equation \ref{eq:ExpandCapacity} gives the steady state temperature
\begin{align}\label{eq:SteadyStateT}
T - T_0 = \frac {c_0 P_\lambda} {\gamma + c_1 P_\lambda}.
\end{align}
Equation \ref{eq:SteadyStateT} behaves as we would expect; the steady state temperature is the ambient temperature in the dark when $P_\lambda =0$.  In the small intensity regime, the $c_1 P^\lambda$ term gives a quadratic correction and in the infinite intensity regime gives $T-T_0 = c_0/c_1$.

The dynamics emerge by from Equation \ref{eq:HeatEq} with the help of Equation \ref{eq:ExpandCapacity}, yielding
\begin{align}\label{eq:SolveHeatEq}
T- T_0 = \frac { c_0 P_\lambda \left(1 - \exp \left[ - \left(\gamma + c_1 P_\lambda \right) t \right] \right) } {\gamma + c_1 P_\lambda } .
\end{align}
Equation \ref{eq:SolveHeatEq} has an onset time constant of
\begin{align}\label{eq:SolveHeatEqTau}
\tau =\frac {1} {\gamma + c_1 P_\lambda} .
\end{align}
When the light is turned off, the material's temperature will decay according to Equation \ref{eq:HeatEq} with $P_\lambda = 0$.  If the light is turned off after an exposure time $t_0$ at power $P_\lambda$, the initial temperature will be given by Equation \ref{eq:SolveHeatEq} at this time, so cooling will yield a time dependence of the form
\begin{align}\label{eq:SolveHeatEq2}
T - T_0 = \frac { c_0 P_\lambda \left(1 - \exp \left[ - \left(\gamma + c_1 P_\lambda \right) t_0 \right] \right) } {\gamma + c_1 P_\lambda } e^{- \gamma (t-t_0) }  .
\end{align}
Clearly, the decay time constant is then given by $\tau = 1/ \gamma$.  In the low power limit, the decay and onset time constants will be the same.

The change in the material's length is proportional to a change in the temperature and the thermal expansion coefficient, so the onset length change is given by
\begin{align}\label{eq:SolveHeatEqForExpandHeat}
\frac {\Delta L} {L} = \alpha(T_0) \frac { c_0 P_\lambda \left(1 - \exp \left[ - \left(\gamma + c_1 P_\lambda \right) t \right] \right) } {\gamma + c_1 P_\lambda }
\end{align}
with steady state length
\begin{align}\label{eq:SteadyStateLength}
\frac {\delta L} {L} = \alpha(T_0) \frac {c_0 P_\lambda} {\gamma + c_1 P_\lambda} ,
\end{align}
and the cooling length change is given by
\begin{align}\label{eq:SolveHeatEqForExpandCool}
\frac {\Delta L} {L} = \alpha(T_0) \frac { c_0 P_\lambda \left(1 - \exp \left[ - \left(\gamma + c_1 P_\lambda \right) t_0 \right] \right) } {\gamma + c_1 P_\lambda } e^{ - \gamma (t-t_0) }  ,
\end{align}
where $\alpha(T_0)$ is the thermal expansion coefficient at temperature $T_0$.  This should not be confused with the conversion rate defined by Equation \ref{eq:ParametersDef}.  Equations \ref{eq:SolveHeatEqForExpandHeat} and \ref{eq:SolveHeatEqForExpandCool} can be used to get the stress response as later described.

We note that the time constant in Equation \ref{eq:SolveHeatEqTau} for pure heating is of the same functional form as that for population conversion between spring types given by Equation \ref{eq:PopTimeConstInfinity}.  Similarly, the onset heating time dependence given by Equation \ref{eq:SolveHeatEqForExpandHeat} and its amplitude given by Equation \ref{eq:SteadyStateLength} Parallels the time dependence given by Equation \ref{eq:PopTimeTurnOn} for the spring-conversion model and its associated steady state population given by Equation \ref{eq:PopTimeInfinity}.

\subsection{Mixed Mechanisms}

Both photo-heating and photo-isomerization have the same temporal form and power dependence.  As such, mechanisms that act individually can be decoupled by fitting the data to a sum of exponentials of varying time constants and amplitudes.  However, the observed universal behavior makes it difficult to determine which process is associated with which time constant unless the model parameters can be individually determined, as we will later describe in the experimental section.  Alternatively, if the mechanisms are coupled -- so that for example the heated material changes the isomer populations that act back on the thermal properties -- separating the mechanisms becomes impossible when using simple additivity.  In these cases the full nonlinear coupled equations need to be solved.  Then, fits of the data to the theory can determine if such mixed processes are at work.  This type of calculation is complex and is left for future work.

\subsection{Bulk Properties from Microscopic Parameters}\label{sec:MicroToMacro}

The above description focuses on the photomorphon, the microscopic building block of the material, while experiments measure the bulk material.  This section relates the microscopic properties to macroscopic measurements, starting with Young's modulus $E$, which connects stress $\sigma$ to strain $u$ according to $\sigma = E u$.   Using Figure \ref{fig:Configurations} to relate the microscopic to macroscopic quantities, we can relate the spring constant of the photomorphon $K$ to Young's modulus,
\begin{align}\label{eq:MacroMicro}
E = K \ell/ A .
\end{align}

The strain response is given by Equation \ref{eq:PopRestLength}
\begin{align}\label{eq:PopDependStrain}
u = \frac {\delta \ell} {\ell} = n_e(t) \cdot \frac { x_b - a } {\ell} ,
\end{align}
where $n_e$ is given by Equations \ref{eq:PopTimeTurnOn} and \ref{eq:PopTimeEvolveTurnOff} during exposure and in the dark.  The strain is thus proportional to the population of photomorphons in their excited state, which varies with exposure and time so can be used to determine the photomechanical response.

Equation \ref{eq:PopDependStrain} is the most general result, and fitting the time-dependent data to this theory determines the various parameters.  Here we use limiting cases to gain an understanding of the big picture.  An obvious limiting case is $t \rightarrow \infty$, which is the photomechanical strain after the system has settled into steady state under light illumination.

Equation \ref{eq:PopDependStrain} in the limit of infinite time to second order in the power is given by
\begin{align}\label{eq:PopDependStrain2ndOrderPower}
u(t \rightarrow \infty) = \frac {\alpha_0} {\beta} \cdot \frac { x_b - a } {\ell} \cdot \left[ I- \frac {\alpha} {\beta} I^2 \right] ,
\end{align}
where we have used Equation \ref{eq:PopTimeTurnOn}.  Thus, the time constant of the time-dependence under illumination yields $\beta + \alpha I$ via Equation \ref{eq:PopTimeTurnOn} and $\beta$ can be determined from the decay time constant according to Equation \ref{eq:PopTimeEvolveTurnOff}.  These two time constants can be used to determine $\alpha$ if the power is known.

In light of the above model, a generalized bulk photomechanical response can be defined by
\begin{align}\label{eq:BulkStrainPM}
u(t \rightarrow \infty) = \sum_{n=0}^\infty \kappa_u^{(n)} I^n \approx \kappa_u^{(1)} I + \kappa_u^{(2)} I^2 ,
\end{align}
where $\kappa_u^{(0)}$ is the pre-strain (strain with no light), and $\kappa_u^{(i)}$ the $i^\text{th}$ photomechanical strain response coefficient.
When the spring model applies, the coefficients are related to each other; for example $\kappa_u^{(2)}/\kappa_u^{(1)} = - \alpha/\beta$.

The exponential time response originates in the population model.  In general, the photomechanical coefficients are given by the response functions $\kappa_u^{(i)} (t; t_1, t_2, \dots, t_i ) $, which relate the intensity at one time to the strain response at another time, or
\begin{align}\label{eq:StrainResponseFunction}
u^{(i)}(t) =  \int dt_1 \int dt_2 \dots & \int dt_i \kappa_u^{(i)} (t; t_1, t_2, \dots, t_i ) \nonumber \\
& \times I(t_1) I(t_2) \dots I(t_i) .
\end{align}
Note that causality is built into the response function so that the intensity precedes the light-induced strain.

The long-time stress when the photomorphon is clamped can be determined from Equation \ref{eq:PopAveForce} by using Equation \ref{eq:PopTimeInfinity} for the population of excited molecules, yielding
\begin{align}\label{eq:StressResponse}
\sigma(t \rightarrow \infty) &= \frac {F(t \rightarrow \infty)} {A} = \frac {k \left( x_b - a \right)} {A} \nonumber \\
& \times \frac {\alpha_0 I} { \beta+ \alpha I  - \alpha_0 I \left(1 - \frac {k} {k^\prime} \right) } .
\end{align}
Writing Equation \ref{eq:StressResponse} in the form
\begin{align}\label{eq:StressResponseBeforeExpand}
\sigma(t \rightarrow \infty) &= \frac {\alpha_0} {\beta} \cdot \frac {k \left( x_b - a \right)} {A} I \nonumber \\
& \times \left( 1 - \left[ \frac{\alpha} {\beta}   - \frac{\alpha_0} {\beta}  \left(1 - \frac {k} {k^\prime} \right) \right] I \right)^{-1}
\end{align}
makes it easy to expand in the limit of small intensity, yielding
\begin{align}\label{eq:StressResponseSmallI}
\sigma(t \rightarrow \infty) &\approx \frac {\alpha_0} {\beta} \cdot \frac {k \left( x_b - a \right)} {A} \nonumber \\
& \times \left( I + \left[ \frac{\alpha} {\beta}  - \frac{\alpha_0} {\beta}  \left(1 - \frac {k} {k^\prime} \right) \right] I^2 \right) .
\end{align}

Equation \ref{eq:StressResponseSmallI} is of the form
\begin{align}\label{eq:BulkStressPM}
\sigma(t \rightarrow \infty) = \sum_{n=0}^\infty \kappa_\sigma^{(n)} I^n \approx \kappa_\sigma^{(1)} I + \kappa_\sigma^{(2)} I^2 ,
\end{align}
where the coefficients $\kappa_\sigma^{(n)}$, which describe the material response, can be determined by direct comparison between Equations \ref{eq:StressResponseSmallI} and \ref{eq:BulkStressPM}.  The most general response function for the stress is of the same form as that of the strain as given by Equation \ref{eq:StrainResponseFunction}.

Consider the special case where the light mostly converts the resting spring into the excited one, and the reverse light-induced process is negligible.  Then, $\alpha = \alpha_0$ and Equation \ref{eq:StressResponseSmallI} becomes
\begin{align}\label{eq:StressResponseSmallIrNoStim}
\sigma(t \rightarrow \infty) \approx \frac {\alpha} {\beta} \cdot \frac {E \left( x_b - a \right)} {\ell} \left( I + \frac{\alpha} {\beta}   \frac {k} {k^\prime}  I^2 \right) ,
\end{align}
where we have expressed the result in terms of Young's modulus.

Comparing Equation \ref{eq:StressResponseSmallIrNoStim} with Equation \ref{eq:BulkStressPM} allows us to relate the measured photomechanical stress response with the microscopic parameters,
\begin{align}\label{eq:StressResponseLinearNoStim}
\kappa_\sigma^{(1)} = \frac {\alpha} {\beta} \cdot \frac {E \left( x_b - a \right)} {\ell}  ,
\end{align}
and
\begin{align}\label{eq:StressResponseQuadNoStim1}
\kappa_\sigma^{(2)}   =  \frac{\alpha} {\beta}   \frac {k} {k^\prime} \cdot \kappa_\sigma^{(1)}.
\end{align}

In the case where the spring constant changes negligibly upon light excitation, or $k \approx k^\prime$, Equation \ref{eq:StressResponseSmallIrNoStim} becomes
\begin{align}\label{eq:StressResponsek=k^prime}
\sigma(t \rightarrow \infty) \approx \frac {\alpha_0} {\beta} \cdot \frac {E \left( x_b - a \right)} {\ell} \left( I + \frac{\alpha} {\beta}  I^2 \right) ,
\end{align}

\begin{align}\label{eq:StressResponsek=k^prime}
\kappa_\sigma^{(1)} = \frac {\alpha_0} {\beta} \cdot \frac {E \left( x_b - a \right)} {\ell}  ,
\end{align}
and
\begin{align}\label{eq:StressResponseQuadk=k^prime}
\kappa_\sigma^{(2)}   =  \frac{\alpha} {\beta} \cdot  \kappa_\sigma^{(1)}.
\end{align}
\subsection{Relationship Between the Stress and Strain Response}\label{sec:StressToStrain}

A comparison of Equations \ref{eq:PopDependStrain2ndOrderPower} and \ref{eq:BulkStrainPM} gives the linear strain response
\begin{align}\label{eq:LinearStrain}
\kappa_u^{(1)} = \frac {\alpha_0} {\beta} \cdot \frac {x_b - a} {\ell} .
\end{align}
Similarly, a comparison between Equations \ref{eq:StressResponseSmallI} and \ref{eq:BulkStressPM} gives the stress response
\begin{align}\label{eq:LinearStress}
\kappa_\sigma^{(1)} = \frac {\alpha_0} {\beta} \cdot \frac {k} {A} \left(x_b - a \right).
\end{align}

Taking the ratio between Equations \ref{eq:LinearStrain} and \ref{eq:LinearStress} gives the relationship between $\kappa_u^{(1)}$ and $\kappa_\sigma^{(1)}$
\begin{align}\label{eq:LinearStressStrainResponse}
\kappa_\sigma^{(1)} = \frac {k \ell} {A} \kappa_u^{(1)} = E \kappa_u^{(1)},
\end{align}
where we have used the definition of Young's modulus given by Equation \ref{eq:MacroMicro} with $K=k$.  Thus, the linear photomechanical response coefficients  $\kappa_u^{(1)}$ and $\kappa_\sigma^{(1)}$ are proportional to each other.  Note that Equation \ref{eq:LinearStressStrainResponse} is a useful relationship that converts between stress and strain response.  As such, a measure of one determines the other one if Young's modulus is known.

\subsection{Photomechanical Efficiency}\label{sec:PMefficiency}

To convert the photomechanical efficiency per photon to the bulk response per volume, we start with Equation \ref{eq:dV-k=k'} for a material that has the same Young's modulus with and without light applied ($k = k^\prime$), yielding
\begin{align}\label{eq:dV-E=E'}
v = \frac {dV} {A \ell} &= \frac {1} {2} \frac {1} {k} \left( F_1 - F_2 \right)^2 /\ell A \nonumber \\ &= \frac {1} {2} \frac {1} {E A/\ell} \left( A \sigma_1 - A \sigma_2 \right)^2 / \ell A,
\end{align}
where we have converted all the microscopic values to bulk ones and expressed the results in terms of strain in lieu of force.  Multiplying out all the lengths and areas, and using the photomechanical stress definition to first-order in the field converts Equation \ref{eq:dV-E=E'} to the form
\begin{align}\label{eq:dV-E=E'2}
v = \frac {1} {2} \frac {1} {E} \left( \kappa_\sigma^{(1)} \right)^2 I^2 .
\end{align}

Equation \ref{eq:dV-E=E'2} is derived from the microscopic picture.  The macroscopic equivalent can be determined from the mechanical energy density stored in a linear strained material with $\sigma = E u$, which yields
\begin{align}\label{eq:BulkEnergy}
\epsilon = \int_0^u du^\prime \sigma(u^\prime) = \int_0^u du^\prime E u^\prime = \frac {1} {2} E u^2 .
\end{align}
For a linear photomechanical response with $u = \kappa_u^{(1)} I$, Equation \ref{eq:BulkEnergy} becomes
\begin{align}\label{eq:BulkEnergyFinal}
\epsilon = \frac {1} {2} E \left( \kappa_u^{(1)}\right)^2 I^2 =  \frac {1} {2} \frac {1} {E} \left( \kappa_\sigma^{(1)}\right)^2 I^2,
\end{align}
which is equivalent to Equation \ref{eq:dV-E=E'2}.

Equations \ref{eq:dV-E=E'2} and \ref{eq:BulkEnergyFinal} suggest an efficiency figure of merit for a material of the form
\begin{align}\label{eq:FOM}
FOM = \frac {1} {E} \left( \kappa_\sigma^{(1)} \right)^2 =  E \left( \kappa_u^{(1)} \right)^2 ,
\end{align}
where the last equality uses Equation \ref{eq:LinearStressStrainResponse} to express the efficiency in terms of the strain response.  Thus, a measure of the clamped stress-photomechanical coefficient and Young's modulus together uniquely defines the figure of merit.

\subsection{Example of a Dye-Doped Polymer}\label{sec:DyeDopedPolyExample}

\begin{figure}\centering
    \includegraphics{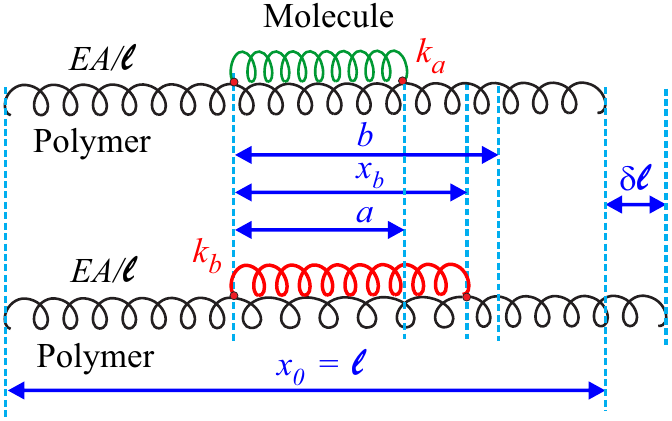}
    \caption{The resting and excited photomorphon for a molecule that is attached attach to a polymer chain by its ends. }\label{fig:PolymerSpringsModel}
\end{figure}

Figure \ref{fig:PolymerSpringsModel}a shows an example of a model of a photomorphon where the ``molecule's"  ends are attached to the polymer, making two springs in parallel that are in turn in series with the rest of the polymer.  The neat polymer is assumed to have a Young's modulus of $E$, so the spring constant of the full polymer chain in the absence of the molecule is $E A / x_0$.

In this example, the microscopic properties of the material can be related to the photomorphon parameters as follows.  First, the segment of the polymer attached to the dye molecule of length $a$ has spring constant
\begin{align}\label{eq:PolySpringConst}
k_e = \frac {E A } {a} ,
\end{align}
making the effective spring constant
\begin{align}\label{eq:EffPolySpringConst}
k_x^{eff} = k_x + \frac {E A } {a} ,
\end{align}
where $k_x=k_a$ for the resting spring representing the molecule and $k_x = k_b$ for the spring constant of the excited molecule.  In the same vein, the spring constant of the polymer ends is given by
\begin{align}\label{eq:PolySpringEnds}
k_p = \frac {E A } {x_0 - a} ,
\end{align}

Using Equations \ref{eq:PolySpringConst} through \ref{eq:PolySpringEnds}, the photomorphon's spring constants are given by
\begin{align}\label{eq:PolySpringConst2}
k =  \frac { EA \left(a k_a + EA \right) } {k_a a \left( x_0 - a \right) + E A x_0} .
\end{align}
and
\begin{align}\label{eq:PolySpringConst3}
k^\prime =  \frac { EA \left(a k_b + EA \right) } {k_b a \left( x_0 - a \right) + E A x_0} .
\end{align}
Putting it all together, the population-weighted elasticity is given by
\begin{align}\label{eq:PopWeightedElasticty}
E & = E_0 \left[1 + \frac {a} {\ell} \cdot \frac {a k_a} {a k_a + E_0 A} \right. \nonumber \\
& \left . +  n_e \cdot \frac {a} {\ell} \cdot \left( \frac {\left( ak_b - a k_a \right) E_0 A} {\left( a k_b + E_0 A \right) \left( a k_a + E_0 A\right)} \right)\right]
\end{align}
where $E_0$ is the neat polymer's elasticity.  As we have shown before, the intensity dependence of $n_e$ -- the excited state population fraction -- is the source of the photomechanical response.  The second term in Equation \ref{eq:PopWeightedElasticty} vanishes if the resting and excited spring constant representing the molecule are the same.

\section{Experiments}

\begin{figure}\centering
    \includegraphics{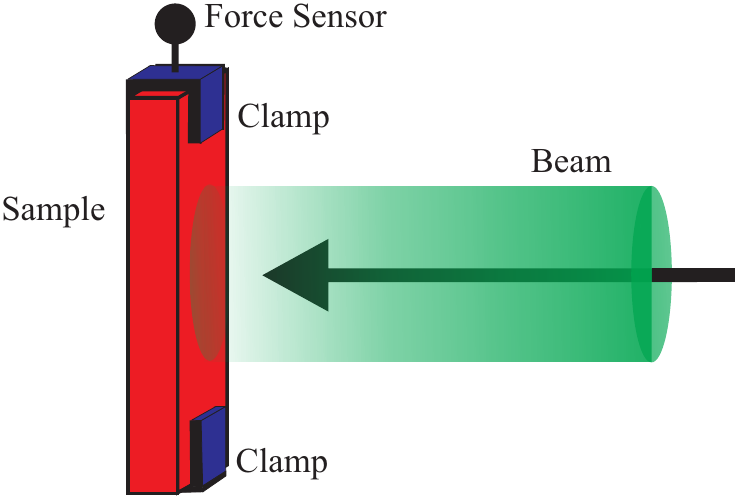}
    \caption{Light illuminates the sample from the side and the vertical component of the stress is measured.  Ideally, the sample is fully illuminated throughout its whole volume.}\label{fig:Experiment}
\end{figure}

The experiments in this paper focus on the clamped configuration, where the force exerted by the sample is measured as a function of time after the light is turned on.  Such an experiment is capable of measuring the loop shown in Figure \ref{fig:IxCycle}.   Figure \ref{fig:Experiment} shows a schematic diagram of the experiment, where light illuminates the sample from the side and the stress is measured in the vertical direction.  The results are best analyzed when the beam is expanded to uniformly illuminate the whole sample, which we implement using a cylindrical lens that forms a line on the sample with a beam waist that is larger than the width of the sample.  Details of the experiment are described in the literature.\cite{bernh18.01}

Microscopic mechanisms acting within the material dependent only on the material's properties.  In contrast, bulk processes are influenced by the geometry of the sample and other experimental conditions. To minimize complications associated with bulk measurements, the sample should be much thinner than the absorption length so that the whole material is uniformly illuminated.  In the other extreme, when most of the light is absorbed near the surface of the sample, the temperature of the dark area will increase through heat diffusion from the bright layer, adding a purely thermal response and making the data more difficult to interpret.\cite{dawso11.02,dawson11.03}  Furthermore, over longer periods of time, the light can bleach the sample, penetrating further over time.  This thick-sample time response has been modelled by Kne\v{z}evi\'{c} and Warner in a liquid crystal elastomer and shown to fit the data.\cite{kneze13.02}  However, when characterizing unknown materials, it is best to eliminate complications by making the experiment as simple as possible.  For this reason, we focus only on an analysis of thin samples with the understanding that thick samples can be modelled with the parameters determined from thin material measurements.

\begin{figure}\centering
    \includegraphics{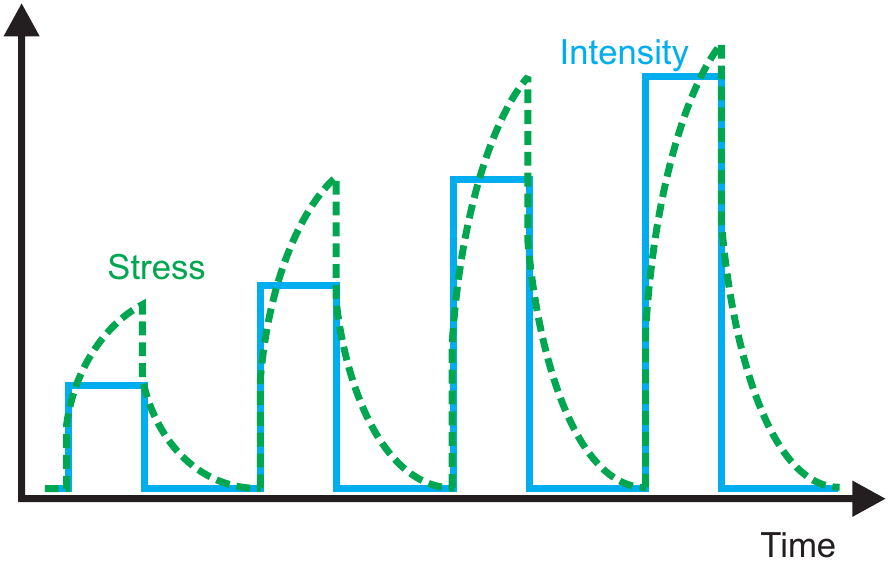}
    \caption{The stress is measured as a function of time as the light is turned on and off with increasing intensity in each cycle.}\label{fig:TimeProfile}
\end{figure}

Figure \ref{fig:TimeProfile} shows a schematic diagram of the time profile of the intensity and stress in a typical experimental run.  The sample is clamped, slightly stressed, then illuminated with a sequence of constant intensity light followed by darkness while the beam is blocked for a time longer than the relaxation time of the material.  In subsequent cycles, the light intensity is increased and the beam unblocked, then blocked.  The cycle is repeated until the full intensity range is probed.  The force on the sensor is continuously recorded at a sampling rate of about 800kHz.  The force is converted to stress by dividing by the sample's cross-sectional area.  The experiment can also be repeated for a range of pre-strains, which can be adjusted with a stepper motor that changes the distance between the clamps for a mounted sample.  All the experiments presented here are for negligibly small pre-strain.

\begin{figure}\centering
    \includegraphics{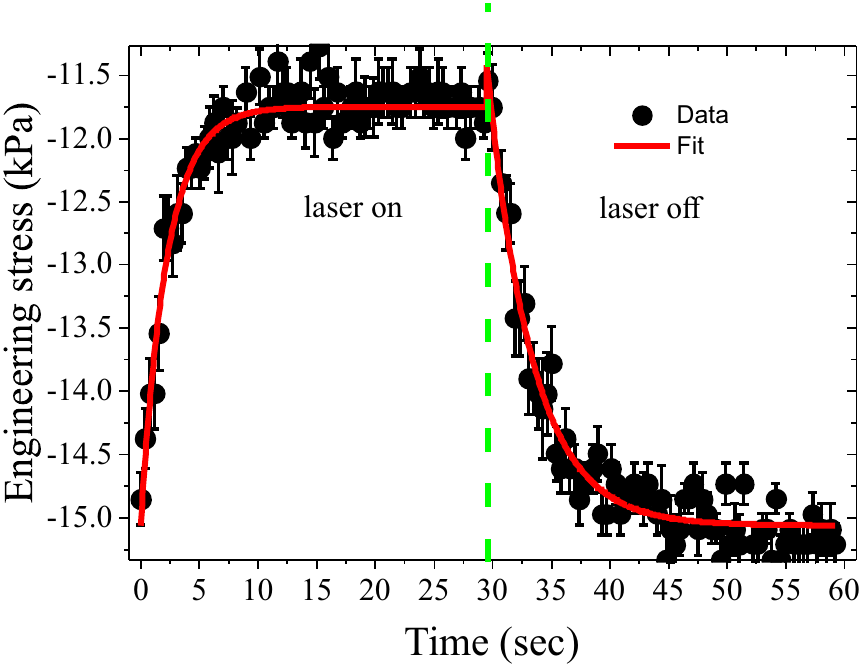}
    \caption{Typical time dependence of the photomechanical stress of a liquid crystal elastomore doped with Disperse Red 1 Azo dye (DO1).}\label{fig:TimeDepend}
\end{figure}

To improve statistics, a smoothing algorithm is applied to the time series and the on-off cycle (when the laser beam is unblocked and block) at a fixed intensity is repeated multiple times.  Figure \ref{fig:TimeDepend} shows an example of the resulting data at one intensity for a DO1-doped liquid crystal elastomer.  The error bars are determined from the standard deviation of the data points that were averaged.  The fit is to a simple exponential both during excitation and relaxation.  Because the data for this sample appear to be well modelled by a single exponential, the amplitudes and time constants of the fits fully characterize the response function.  If the results require a multiple exponential fit, as will be the case in other samples, then more parameters are required to reconstruct the time-dependence.

The uncertainty for the parameters is determined from the fitting routine, which is a measure of how well the theory fits the data.  Another uncertainty -- which we calculate from the standard deviation of the parameters given by fitting over several repetitions of the measurement -- quantifies the reproducibility of the measurement.  Finally, we repeat measurements on samples carved out of the same parent material and find that the parameters can vary by as much as 10\% even though the samples should be identical.  This variation originates in both the material and in mounting the samples, which can result in slight variations in the sample angle, strength of clamping, etc.  As such, if one is interested in comparing the response functions of different materials, a 10\% uncertainty must be assigned since the uncertainties of the measurements for a fixed sample in the setup is much smaller than variations between experiments on different samples prepared from the same stock.

\subsection{Force Sensor Corrections}

\begin{figure}\centering
    \includegraphics{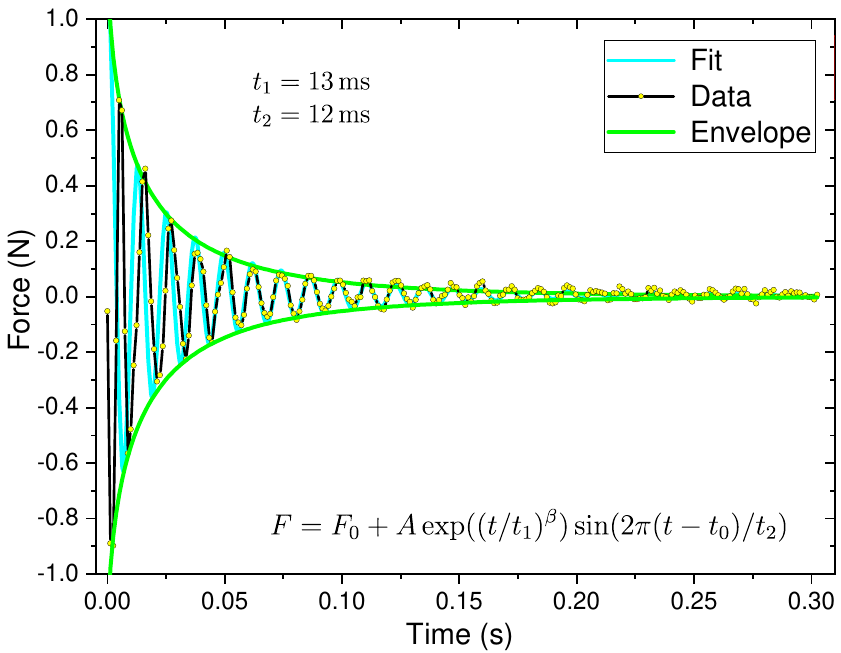}
    \caption{The response of the stress sensor after an applied impulse, a fit to a damped stretched oscillator and the envelope function.}\label{fig:CalibratePlot}
\end{figure}

The ideal force sensor has an infinite effective spring constant and an instantaneous response.  Differences from the ideal are immaterial if the sample's effective spring constant is much smaller than the sensor's spring constant and if the response time of the material is much longer than that of the sensor.  Figure \ref{fig:CalibratePlot} shows the response of the sensor after the applied force is turned off and a fit to the theory of a damped stretched oscillator.

A damped harmonic oscillator decays as a sinusoidal function with an exponentially decaying amplitude.  The force senor, being of a more complex geometry, deviates from this type of response.  We model the envelope as a stretched exponential to account for the longer tail, yielding the time dependence
\begin{align}\label{eq:ModelSensor}
F = F_0 + a \exp((t/t_1)^\beta) \sin (2 \pi (t - t_0)/t_2) .
\end{align}
The parameter $\beta$ is a measure of the ``stretch" of the exponent.  The initial part of the data is slightly chirped so that the oscillations fit well only after the first couple oscillations.  However, the time constant $t_1 = 13 \, \text{ms}$ is much shorter than the response of the material, so that the detector can be assumed to respond instantaneously.  A faster response can be treated by deconvoluting the data with the experimentally-determined response function of the sensor.

\begin{figure}\centering
    \includegraphics{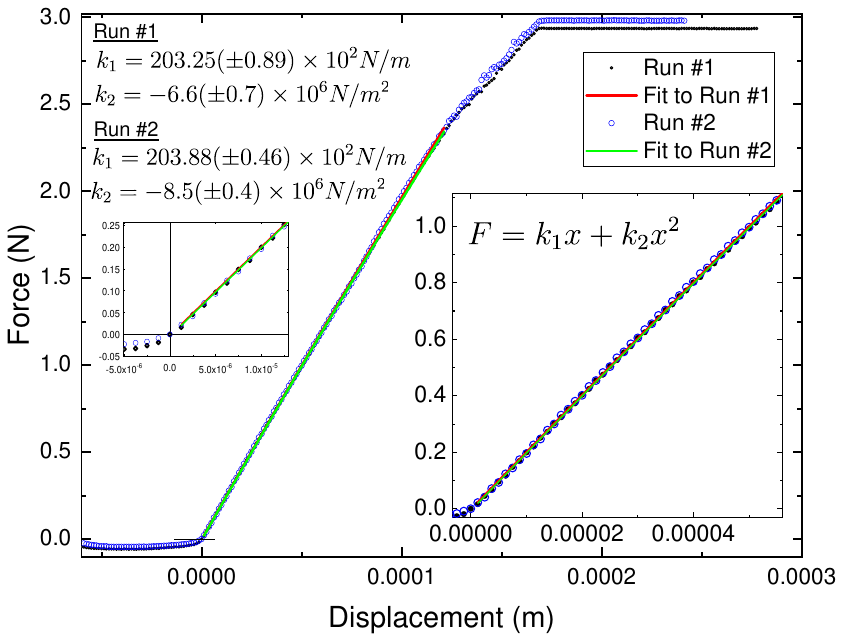}
    \caption{The force sensor reading as a function of displacement for two separate runs.  The quadratic fit to the data is used for calibration.  The right inset shows a magnified view of the small displacement limit and the left inset is zoomed in more to show that the curve is linear to the origin.}\label{fig:Calibrate}
\end{figure}

To take into account the effect of the sensor's finite spring constant, we start by measuring it when the two clamps are in direct contact, then reading the force from the sensor as a translation stage pushes the bottom clamp into the top one.  The measured force  as a function of displacement of the sensor (read from the translation stage), is used as a calibration function of the sensor.  From this calibration, a measure of the force also gives the displacement of the upper clamp.  Note that the translation stages starts moving prior to it coming in contact with the sensor mount, hence the flat region.  The origin is set to the point where contact is made.

Figure \ref{fig:Calibrate} shows two separate runs and a fit to a quadratic function.  In the small displacement limit (for strains less than $10^{-4}$), the curve is highly linear, the spring constants of two runs agree within experimental uncertainty and differ by less than 1\%.  Thus, as the data will show, the uncertainty in the force reading is much smaller than fluctuations in the data due to acoustical noise and the time constant of the response is shorter than the time scales of the processes measured.

\subsubsection{Young's Modulus Measurement}

\begin{figure}\centering
    \includegraphics{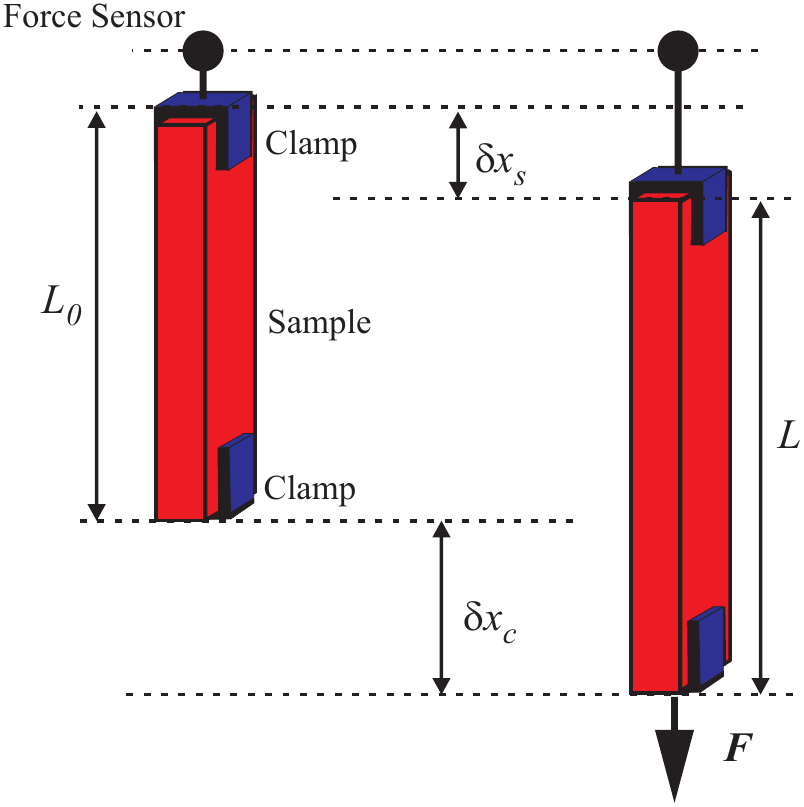}
    \caption{The sample and the sensor both deform when a force is applied to the bottom of the sample.}\label{fig:SensorSample}
\end{figure}

Viewing the force sensor as a spring of force constant $k_1$ in series with the sample, of spring constant $\kappa = E A/L$, we need to determine $\kappa$ given a measurement of the force as a function of the displacement of the end of the sample.  If the displacement of the end of the sample as measured by the bottom clamp on the translation stage is $\delta x_c$ and the measured force by the sensor is $F$, the sensor adds $\delta x_s$ to the displacement in the amount $x_s = F/k_1$, as shown in Figure \ref{fig:SensorSample}.  As such, the length change of the material $\delta x_m$ is given by
\begin{align}\label{eq:LengthChangeAccounting}
\delta x_m = L-L_0 = \delta x_c - \delta x_s = \delta x_c - F/k_1.
\end{align}
Then, the spring constant of the sample is given by
\begin{align}\label{eq:SampleSpringConst}
\kappa = F/ \delta x_m = \frac {F} { \delta x_c - F/k_1},
\end{align}
where we have used Equation \ref{eq:LengthChangeAccounting}.  Note that when $k_1 \rightarrow \infty$, the sensor does not deform, and the spring constant of the sample is given simply by the ratio of the measured force to the displacement of the end.  Young's modulus is then simply given by $E = \kappa L/A$.

\subsubsection{Thermal Expansion Coefficient}

The clamped configuration can be used to measure the thermal expansion coefficient, which is a useful parameter for modeling the mechanisms of the photomechanical response.  The thermal expansion coefficients $\alpha(T)$ is defined by
\begin{align}\label{eq:ThermalExpand}
\frac {\delta L} {L} = \alpha(T) \delta T,
\end{align}
where $\alpha(T)$ can depend on temperature.  Note that negatve $\alpha$ describes a material that contracts with increased temperature.  For an unclamped sample, $\delta L$ and $L$ are measured when the temperature changes by a sufficiently small amount of $\delta T$ such that $\alpha(T)$ is approximately constant over the temperature interval $\delta T$.

Consider first an ideal force sensor with infinite spring constant.  Once the sample length changes, it can be squished by an amount $- \delta L$ to restore it back to length $L$ by a force
\begin{align}\label{eq:ThermalExpandForce}
- \delta F = \frac {E(T+\delta T) A} {L+\delta L} (- \delta L),
\end{align}
where we must use Young's modulus of the material at the elevated temperature.  We note that compression corresponds to a negative force, thus we define {\em  $F<0$ when the sample pushes up on the sensor and $F>0$ when it pulls the sensor downward.}

Eliminating $\delta L / L $ from Equations \ref{eq:ThermalExpand} and \ref{eq:ThermalExpandForce} yields in the limit of small $\delta T$
\begin{align}\label{eq:ThermalExpandClamped}
\lim_{\delta T \rightarrow 0} \left[ \frac { \delta F} {\delta T}=  \frac {\left( E(T) + \frac {\partial E} {\partial T} \delta T \right)A \alpha} {1 + \alpha \delta T} \right]  \rightarrow  \frac {\partial F} {\partial T} = E(T) A \alpha,
\end{align}
which gives $\alpha$ in the form
\begin{align}\label{eq:ThermalExpandClamped2}
\alpha(T) = \frac {\partial F} {\partial T} \frac {1} { E(T)  A},
\end{align}
where $\partial F / \partial T$ is measured and $E$ is the known Young's modulus of the material, which is separately measured.

To prevent buckling, the sample is slightly stretched in the holder so that photomechanical expansion never increases the length beyond the distance between the clamps.  A force $F(T)$ at temperature $T$ is applied at the start of the experiment for such pre-stretching.  Then, the measured force upon temperature increase is due to material stress as well as the the change in the sample's length due to a deformation of the sensor.  In the pre-stressed configuration, the force read by the sensor is
\begin{align}\label{eq:PressStressForce}
F(T) = + \frac {E(T) A} {L_0 (T)} \Big( L(T) - L_0(T) \Big),
\end{align}
where $L_0$ is the resting length of the material, $L$ the stretched length, and the positive sign emphasizes that the sample is being stretched.  Note that the area will also change with temperature, but $\delta A \delta L \ll A \delta L$, so we ignore it.  Up to this point, we explicitly labeled all the quantities that depend on temperature.  To reduce clutter, the temperature dependence will be implicitly assumed in what follows.

The change in the force when the temperature increases can be calculated by taking the temperature derivative of Equation \ref{eq:PressStressForce}, yielding
\begin{align}\label{eq:PressStressForceDerivative}
\frac {\partial F} {\partial T} &= \frac {A} {L_0} \left[ \left( \frac {\partial E} {\partial T} - \frac {E} {L_0} \frac {\partial L_0} {\partial T} \right) \left(L - L_0 \right) \right. \nonumber \\
& + \left. E \left( \frac { \partial L} {\partial T} - \frac {\partial L_0}  {\partial T} \right) \right] .
\end{align}
Note that for the ideal sensor, $\partial L / \partial T$ vanishes because the sample length will remain unchanged, but the resting length $L_0$ changes from thermal expansion.  Assuming that the sensor and the sample are in series and that the ends of the sensor/sample system are constrained, the total length must be constant so the length change of the sensor plus the length change of the material must add to zero, hence $\delta L + \delta F/ k_1 = 0$, and
\begin{align}\label{eq:SampleDisplacement}
\frac { \partial L} {\partial T} = - \frac {1} {k_1} \frac { \partial F} {\partial T} .
\end{align}

Using Equations \ref{eq:PressStressForceDerivative} and \ref{eq:SampleDisplacement} and solving for the thermal expansion coefficient $(1/L_0) \partial L_0 / \partial T = \alpha$ yields
\begin{align}\label{eq:ThermalExpandCoefWithSensor}
\alpha = \frac {1} {E} \frac {\partial E} {\partial T} \left( 1 - \frac {L_0} {L} \right)  + \frac {1} {L} \frac {\partial F} {\partial T} \left( - \frac {1} {k_1} - \frac {L_0} {E A}\right).
\end{align}
Equation \ref{eq:ThermalExpandCoefWithSensor} can be re-expressed in a more convenient form using Equation \ref{eq:PressStressForce}, yielding
\begin{align}\label{eq:ThermalExpandCoefWithSensorF}
\alpha = \frac {1} {L} \left[ \frac {1} {E} \frac {\partial E} {\partial T} \left( \frac {F L_0} {EA} \right)  - \frac {\partial F} {\partial T} \left(  \frac {1} {k_1} + \frac {L_0} {E A}\right) \right] .
\end{align}
Note that when $k_1 \rightarrow \infty$, the sensor does not deform and $L=L_0$.  Then, Equation \ref{eq:ThermalExpandCoefWithSensorF} with no force initially applied ($F=0$) reduces to Equation \ref{eq:ThermalExpandClamped2}, as we expect.

\section{Results Discussion}

This section starts with a discussion of the mechanisms of the photomechanical response in dye-doped PMMA polymer and how they can be determined by comparing the response of DR1 dopants -- which isomerize, and DO11 -- which don't.  The dopant molecules are shown in Figure \ref{fig:molecules}.  Dye-doped polymers are found to be dominated by photothermal heating followed by thermal expansion with only a small contribution from light-induced photo-isomerization.  Then, measurements of polydomain liquid crystal elastomers are compared with direct measurements of the sample's temperature change to remove the thermal contribution.  The measured parameters are used to deduce the microscopic properties of the photomorphon, showing that the parameters so determined agree with expectations based on estimates of molecular length changes associated with the shape change during photoisomerization.  Then, the efficiency of several materials are reviewed and compared using the bulk figure of merit defined by Equation \ref{eq:FOM}.

\subsection{Mechanisms}\label{sec:Mechanisms}

\subsubsection{Dye-Doped PMMA Polymer}

\begin{figure}\centering
    \includegraphics[scale=0.7]{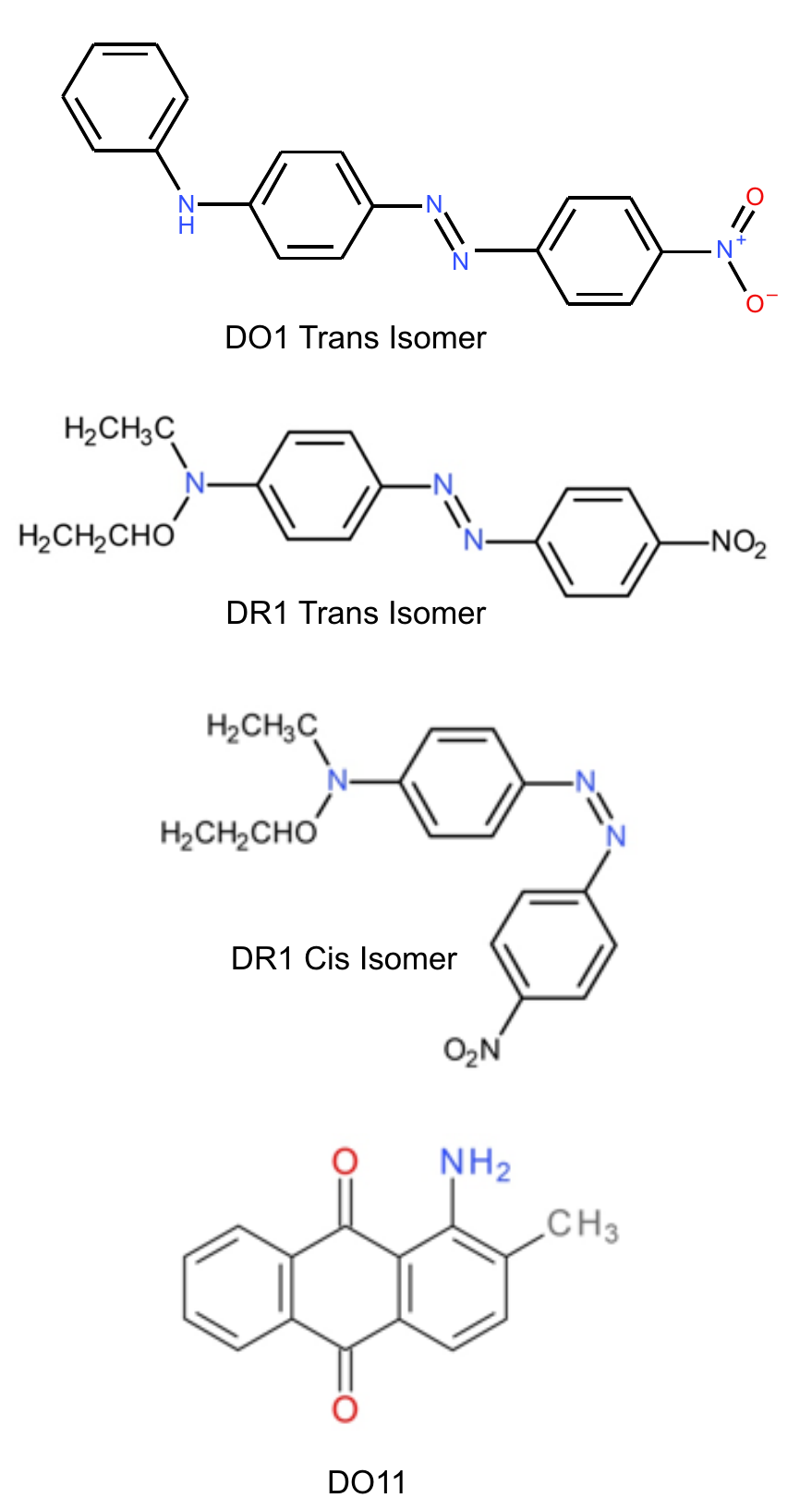}
    \caption{Dopant dyes.}\label{fig:molecules}
\end{figure}

Figure \ref{fig:Dr1DO11} shows the amplitude of photomechanically-induced stress, determined from fits similar to the one shown in Figure \ref{fig:TimeDepend}, as a function of intensity for DR1 dye doped in poly (methyl mehacrylate) (PMMA) polymer and DO11-doped PMMA.  Two polarizations are shown -- one along the length of the fiber and the other perpendicular to it.  The fits are to a second-order polynomial under the constraint that the stress must vanish at zero intensity.  The error bars reflect our estimate of 10\% uncertainty as described above.

Given the anisotropy of the photo-isomerization process, the polarization dependence of the induced stress provides data that can be used to separate the mechanisms of orientational hole burning and photothermal heating.  Orientational hole burning is the process by which light is absorbed by a molecule, changing its shape through photoisomerization, and leading to a population of molecules that is oriented away form the light's polarization.\cite{sekka91.02,dumon92.02,sekka92.03,dumon94.01} Figure \ref{fig:molecules} show the trans isomer, which is the lowest energy state of the DR1 molecules, and the Cis isomer, which is the shape after photoexcitation.  These two forms can be modelled as springs of differing equilibrium lengths and force constants.

During prolonged exposure to polarized light, the molecule oriented along the light's polarization axis are excited, then change shape.  The net result is a decrease in length of the material along the light's polarization axis.  Upon relaxation from the cis to trans state, a random thermal process, the resulting trans molecule randomly re-orient into an isotropic distribution.  Consequently, molecules along the polarization axis are depleted, resulting in a ``hole" in the orientational distribution function at the angle of the light's polarization.  As a result, the material shrinks in the direction of polarization as the rod-like molecules are depleted in that direction and expands in the perpendicular direction as the population of molecules grow away form the light polarization axis.  This orientational state is long lived because the longer trans molecule is entangled in polymer chains.

Alternatively, the molecules might not be strongly coupled to the polymer, so that the shape change at the molecular scale does not change the bulk polymer's shape.  Since photon absorption leads to heat being deposited in the polymer, it will change dimensions due to thermal expansion, leading to an isotropic strain if the material is isotropic, as the PMMA polymers in our studies are usually prepared to be.  Liquid crystal elastomers, on the other hand, can be prepared to be globally isotropic or oriented, so heating can lead to anisotropic expansion/contraction.

Consider first an isotropic material.  The degree of heating, then, is independent of the light's polarization, leading to thermal expansion with strain that is independent of the polarization.  Angular hole burning, on the other hand, will lead to material expansion perpendicular to the light's polarization -- thus adding to the transverse strain -- while acting in opposition to thermal expansion along the polarization direction and decreasing the strain.  The induced stress will show the same behavior, so the difference in light-induced stress along and perpendicular to the long length of the sample is indicative of a contribution of angular hole burning.

Molecules that do not change shape, on the other hand, will not result in angular hole burning and can act as a control.  In our experiments, we use the dye DR1, which is well known to photoisomerize, and DO11, which does not.  To ensure that the degree of heating is the same in both samples at the measurement wavelength, we adjust the concentrations in inverse proportion to the absorbance.  At the measurement wavelength of $\lambda = 488 \, nm$, the ratio of DO11 to DR1 dye in PMMA polymer must be 2.76 to 1 by weight to make the optical absorbance the same.

\begin{figure}\centering
    \includegraphics{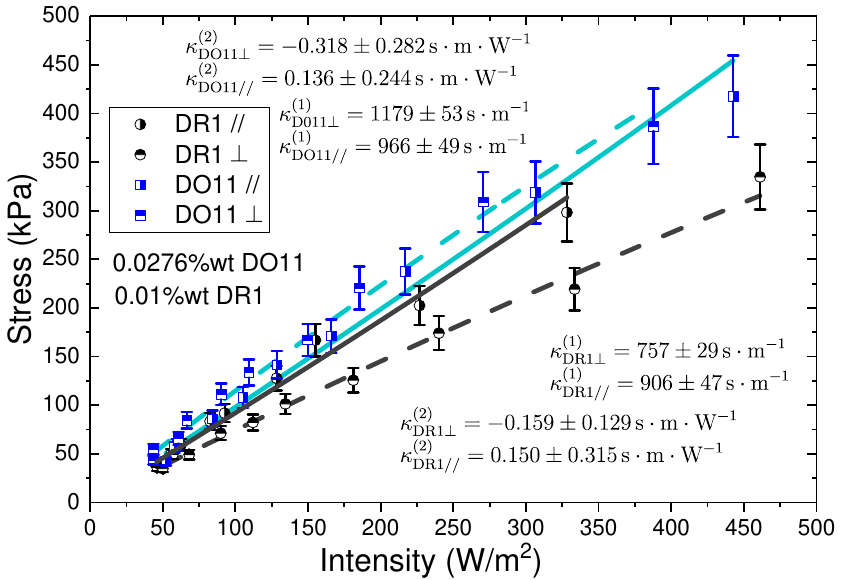}
    \caption{Intensity dependence of photomechanical Stress of DR1-doped PMMA and DO11-doped PMMA for excitation polarizations along and perpendicular to the fiber.}\label{fig:DR1DO11}
\end{figure}

Figure \ref{fig:DR1DO11} shows that the linear photomechanical response $\kappa_\sigma^{(1)}$ is the same within experimental uncertainties for both polarizations in DO11, implying that there is no angular hole burning, as expected for a molecule that does not change shape under light exposure.  However, for DR1, the expansion induced by the perpendicular polarization is systematically lower than the parallel one, the opposite of what we would expect for angular hole burning, thus ruling it out.

The physical and optical properties of a polymer are known to depend on the material's thermal and mechanical history, which can affect the material at various stages of processing.  For example, the drawing process used to make the fiber\cite{kuzyk91.01} can leave it in a non-equilibrium state where the chains are aligned along its length.\cite{kuzyk06.06}  This effect can be minimized during drawing, but not eliminated, by pulling the fiber slowly under low stress.  The magnitude of the effect can be measured by annealing the fiber and measuring the length change.

\begin{figure}\centering
    \includegraphics{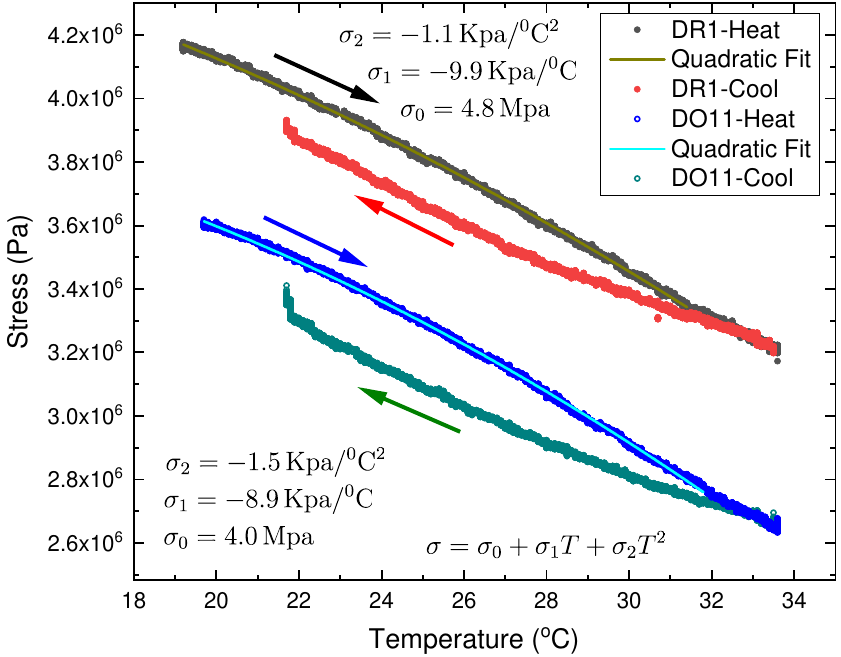}
    \caption{Stress as a function of temperature for DR1- and DO11-doped PMMA as the clamped sample is heated and then cooled.}\label{fig:Stress-Temp}
\end{figure}

Both DR1/PMMA and DO11/PMMA fibers were annealed for a week at 94$^o$C prior to measurements to remove residual stress.  Stress relaxation was confirmed by the observations that both fibers contracted by 4-5\%.  Figure \ref{fig:Stress-Temp} shows a plot of the stress as a function of temperature of two pre-annealed fibers clamped to a stress sensor.  The fiber is initially slightly stretched in the holder to prevent buckling when it expands.  The stress is observed to decrease as the temperature is increased and the fiber expands, then decreases when the sample is cooled.

The observed hysteresis implies that the equilibrium length of the fiber has decreased as a result of the heating/cooling cycle.  It is likely that the increased length of the fiber after it is cooled is due to the initial applied stress when the polymer softens at elevated temperature.  When measuring the photomechanical response, the parallel and perpendicular polarizations shown in Figure \ref{fig:DR1DO11} are measured in sequence.  Thus, the observed difference in the response might be due to the sample's thermal history from laser heating.  As such, these results are most likely not an indication of orientational hole burning.

We can average the perpendicular and parallel response of each material to get the average photomechanical constants.  For DR1 and DO11 we get $\kappa_\text{DR1}^{(1)} = 830 s/m$ and $\kappa_\text{DO11}^{(1)} = 1070 s/m$.  A quadratic fit to the Stress-Temperature data shown in Figure \ref{fig:Stress-Temp} reveals that the response is slightly different between DO11 and DR1.  Using the fit parameters in Figure \ref{fig:Stress-Temp}, we can calculate the linear stress response using
\begin{align}\label{eq:LinearTempStress}
\frac {d \sigma} {dT} = \sigma_1 + 2 \sigma_2 T ,
\end{align}
which yields at $T= 20^o \text{C}$
\begin{align}\label{eq:LinearTempStressResult}
\frac {d \sigma } {dT}_{DR1}  =  5.4 \times 10^4 \text{Pa} / ^o \text{C}
\end{align}
and
\begin{align}\label{eq:LinearTempStressResult2}
\frac {d \sigma } {dT}_{DO11} &=  6.8 \times 10^4 \text{Pa} / ^o \text{C}.
\end{align}

If the photomechanical response originates in photothermal heating, the ratio of the two material's linear response would be the same as the temperature-dependent stress.  The stress ratio and the photo mechanical response ratios are both 1.3, showing that the response is most likely fully due to thermal expansion.  Furthermore, an increase in temperature should increase the cis population of DR1.  Since the cis molecules are smaller, this effect opposes thermal expansion of the polymer.  As such, one would expect DR1 to have a smaller temperature-dependent stress, as is observed.  We emphasize that these are ballpark numbers and that variability of materials might be comparable to these differences.

Table \ref{tab:results} summarizes the results.  The quadratic response determined from the fits is of the same order of magnitude as the uncertainty, suggesting that the curvature for dye-doped PMMA is negligible and the data is consistent with approximate linearity of the stress as a function of intensity.  As we later discuss, elastomeric materials have a large nonlinear term, but the uncertainties are also generally large.

\begin{figure}\centering
    \includegraphics{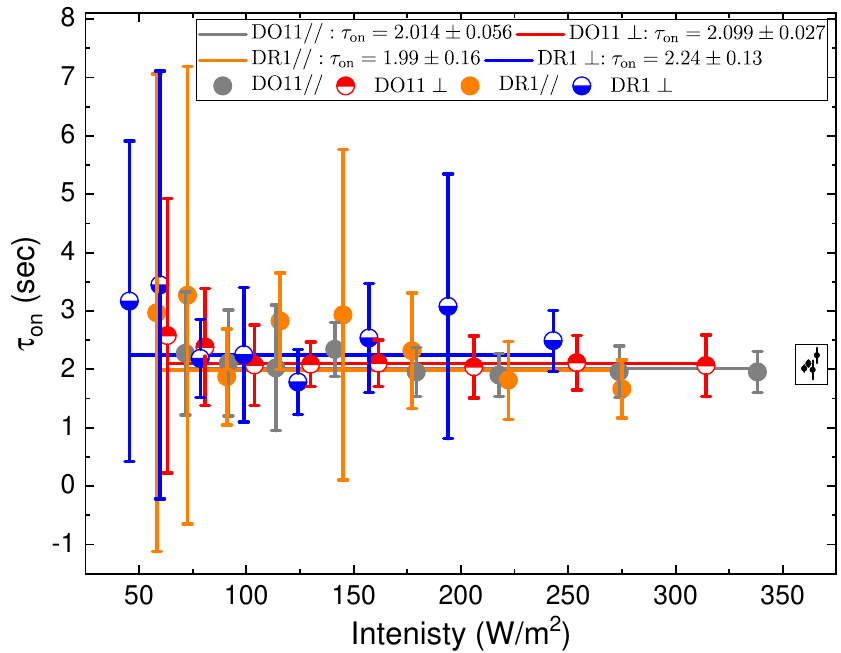}
    \caption{The time constants of the turn on of stress upon illumination for DR1/PMMA and DO11/PMMA for both polarizations.  The tiny inset on the right plots the average time constant for each sample and polarization as well as the error bars derived from the fits.}\label{fig:TimeConstants}
\end{figure}

\begin{table*}
\begin{tabular}{c c c c c c c}
  \hline
   Material & $\kappa_\sigma^{(1)} (s/m)$ & $\kappa_\sigma^{(2)} (s \cdot m / W^2)$ & E(Mpa) & $\kappa_u^{(1)} = \kappa_\sigma^{(1)} /E (s \cdot m/N$) & $\tau_\text{on}$ (s) & FOM ($10^{-4} s^2 / N $)\\ \hline
  Monodomain LCE & -23.4 $\pm$ 1.2 & 8.5 $\pm$ 10.5 & 2.99 $\pm$ 0.02 & 7.8 $\pm$ 0.4 & & 1.83 $\pm$ 0.13\\
  Polydomain LCE & -4.5 $\pm$ 1.9 & -30 $\pm$ 5 & 1.26 $\pm$ 0.01 & 3.6 $\pm$ 1.5 & $\approx 5s$ for $I \rightarrow 0$ & 0.16 $\pm$ 0.10\\
  CNT LCE & -0.98 $\pm$ 0.07 & -5 $\pm$ 1 & 0.457 $\pm$ 0.0002 & 2.1 $\pm$ 0.2 & & 0.020 $\pm$ 0.002\\
  DR1-PMMA; // & 906 $\pm$ 47 & 0.150 $\pm$ 0.315 & 2240 $\pm$ 200 & 0.37 $\pm$ 0.04 & 1.99 $\pm$ 0.16 & 3.66 $\pm$ 0.42 \\
  DR1-PMMA; $\perp$ & 757 $\pm$ 29 & -0.159 $\pm$ 0.129 & 2240 $\pm$ 200 & 0.44 $\pm$ 0.04 &  2.24 $\pm$ 0.13 & 2.56 $\pm$ 0.30\\
  DO11-PMMA; // & 966 $\pm$ 49 & 0.136 $\pm$ 0.244 & 2240 $\pm$ 200 & 0.17 $\pm$ 0.02 & 2.014 $\pm$ 0.056 & 4.16 $\pm$ 0.48\\
  DO11-PMMA; $\perp$ & 1179 $\pm$ 53 & -0.318 $\pm$ 0.282 & 2240 $\pm$ 200 & 0.18$\pm$ 0.02 & 2.099 $\pm$ 0.027 & 6.20 $\pm$ 0.72\\
  \hline
\end{tabular}\caption{A summary of the measured photomechanical response, Young's modulus and figure of merit for all measured materials.\label{tab:results}}
\end{table*}

Figure \ref{fig:TimeConstants} shows the measured time constants associated with the data in Figure \ref{fig:DR1DO11} from the photomechanical stress as determined from the time dependence of the data that typically is of the form shown in Figure \ref{fig:TimeDepend}.  The DR1 dye and DO11 dye-doped PMMA samples where prepared with concentrations that yield the same amount of light absorption at the measurement wavelength.  The tiny inset to the right shows a summary of the time constants of the two samples and two polarizations.  The error bars of all the time constants overlap, except for that of DR1$_{//}$ and DO11$_\perp$, which are close enough to conclude that the time constants are the same within statistical fluctuations.  This strongly suggests that the mechanism of the response is the same in both samples, implying that the photothermal response is the dominant one.

\subsubsection{Dye-Doped Liquid Crystal Elastomers}

Next we apply the clamped configuration to characterize a liquid crystal elastomer doped with Disperse Orange dye (DO1, shown in Figure \ref{fig:molecules}).  The elastomer was soaked in a 0.1 percent by weight solution of DO1 in toluene to infuse it into the neat elastomer.  The length of the poly-domain sample between the clamps is $4.25\, \text{mm}$ and the dimensions of the approximately rectangular cross-section is $2.014\, \text{mm} \times 0.44\, \text{mm}$.  $0.40 \, \text{mm}$ is thick enough to absorb all of the light from the $\lambda =488\, \text{nm}$ line of an Argon/Krypton laser, which illuminates the whole face of the sample.  A shutter turns the light on and off over 120 second cycles.

\begin{figure}\centering
    \includegraphics{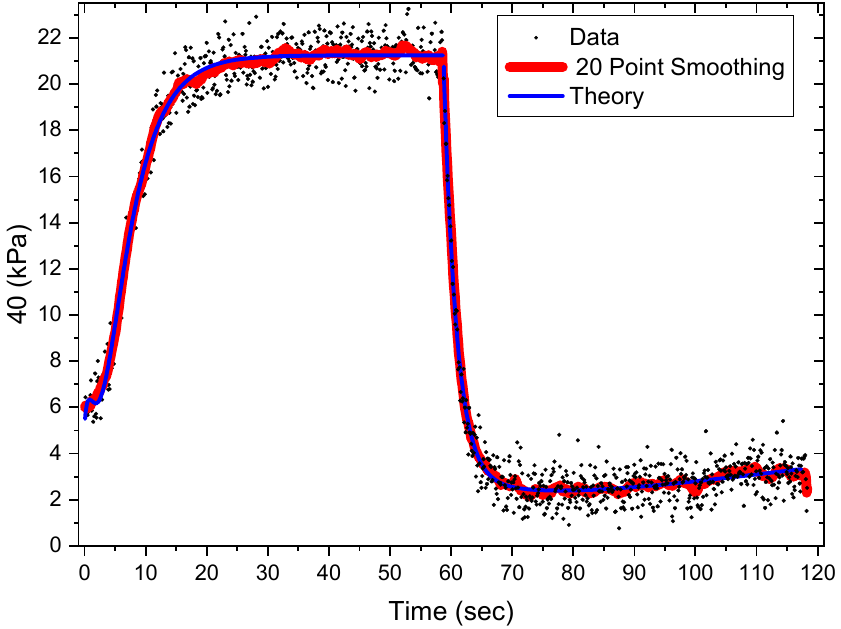}
    \caption{Typical response for a polydomain liquid crystal elastomer sample (points), twenty-point smoothing (thick red points) and triple exponential fit.}\label{fig:TypicalPolyDomain}
\end{figure}

Figure \ref{fig:TypicalPolyDomain} shows a typical run.  Though the experiment is built on a vibration isolated optical table, the observed fluctuations of the data originates in ambient sound and air currents that excite the elastomer.  Without a sample, the stress sensor noise is much smaller, as shown in Figure \ref{fig:Calibrate}.  Twenty-point smoothing of the data shows the expected exponential behavior.  A good fit to the data requires a triple exponential.

The turn-on response immediately after the laser exposes the sample is modelled by
\begin{align}
    \sigma & =
    A_{0} + A_{1} \left(1-\exp\left(-t/t_{1}\right)\right) \nonumber \\
     & + A_{2} \left(1-\exp\left(-t/t_{2}\right)\right) + A_{3} \left(1-\exp\left(-t/t_{3}\right)\right) .
\end{align}
When the laser is turned off after an exposure time of $t_0$, the function used is
\begin{align}
    \sigma &=   A_{0} + A_{1} \left(1-\exp\left(-t_0/t_{1}\right)\right) \left(\exp\left(-\left(t-t_{0}\right)/t_{4}\right)\right) \nonumber \\
    & + A_{2} \left(1-\exp\left(-60/t_{2}\right)\right) \left(\exp\left(-\left(t-t_{0}\right)/t_{5}\right)\right) \nonumber \\
    & + A_{3} \left(1-\exp\left(-60/t_{3}\right)\right) \left(\exp\left(-\left(t-t_{0}\right)/t_{6}\right)\right)
\end{align}
The one/off data is fit simultaneously to get the full set of parameters including $t_0$, the time when the beam is turned off.  These multiple exponentials allow for multiple mechanisms as well as possible instrumental artifacts, with both the photo-thermal heating mechanism (Equation \ref{eq:PopTimeTurnOn}) and photo-isomerization mechanism (Equation \ref{eq:SolveHeatEqForExpandHeat}) both having these exponential forms.  Note that $t_0$ given by the fitting procedure varies negligibly from $t_0 = 60 \, s$, the time the laser is on.

\begin{figure}\centering
    \includegraphics{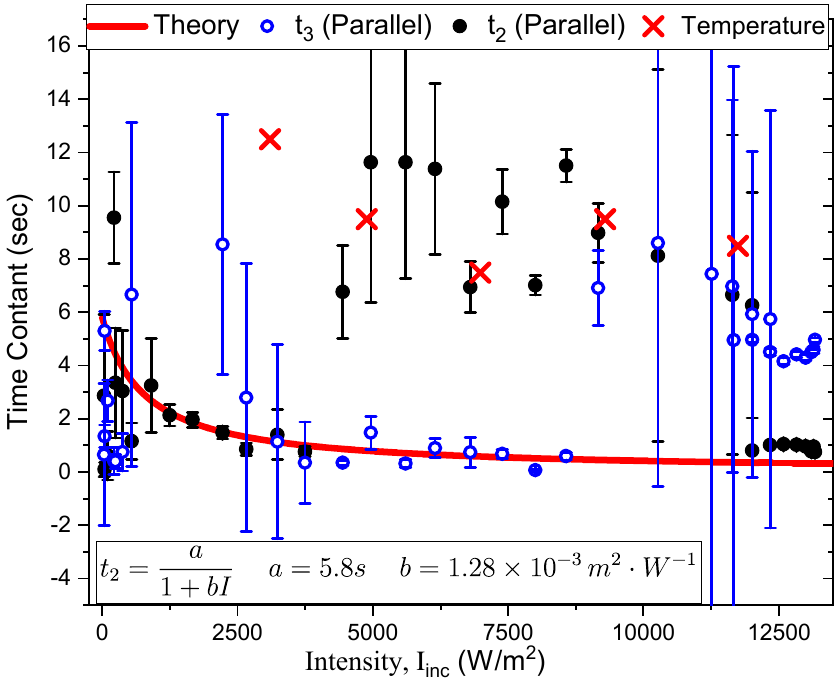}
    \caption{The time constants determined form a fit of the time-dependent stress in a thick liquid crystal elastomer doped with DO1 dye (points) and the theory of an onset process (cruve). The red crosses show the time constant of the photothermal heating response, determined directly from the temperature measurement shown in Figure \ref{fig:TempRaw}.}\label{fig:TimtConstant2and3}
\end{figure}

The experiment is repeated for a range of intensities from $30\, \text{W} \cdot \text{m}^{-2}$ to $1.3 \times 10^4\, \text{W} \cdot \text{m}^{-2}$.  At each intensity, the five time constants $t_i$ and three amplitudes $A_i$ are determined.  Figure \ref{fig:TimtConstant2and3} shows the time constants $t_2$ and $t_3$, which we claim represent the onset time for the photo-isomerization mechanism of the stress response, for which the evidence follows.

We first digress to show how the dependence of the time constants on the intensity illustrates an important issue that can arise in the fitting process, and that is the ambiguity as to which parameter -- such as time constant -- is associate with which process.  The solid black points and the open blue points are determined form the fitting routine, which chooses which label (i.e.  $t_2$ or $t_3$) to use at each intensity.  A plot of the two time constants together suggests that the data might need to be stitched together to get a continuous curve. The red curve is a fit to the black data points representing $t_2$ in the low-intensity range, which is observed to pass through the blue points at higher intensities.  This suggests that the labels $t_2$ and $t_3$ need to be interchanged in the different regions.

The turn-on time constant data is fit to the curve
\begin{align}\label{eq:timeConstForm}
t_2 = \frac {a} {1+bI} ,
\end{align}
the form predicted for both photo-isomerization (Equation \ref{eq:PopTimeConstInfinity}) and photothermal heating (Equation \ref{eq:SolveHeatEqTau}).  Note that we will focus only on the lower intensity regime that is well within the optical damage threshold for the sample.  The fit curve generated from the data below $4,000\, \text{W} \cdot \text{m}^{-2}$ fits a locus of data throughout the full intensity range.

\begin{figure}\centering
    \includegraphics{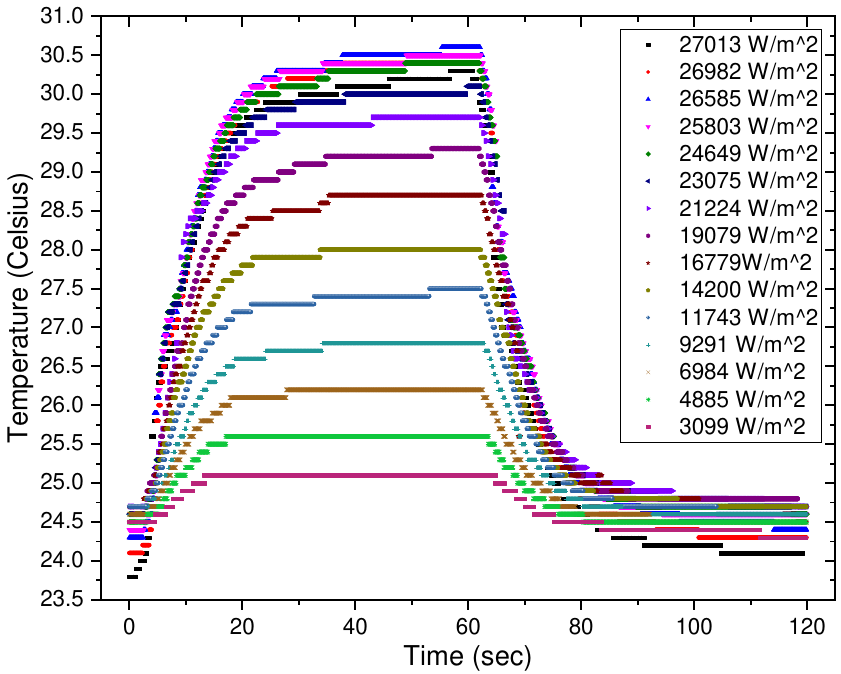}
    \caption{The temperature measured at the back of the DO1-doped elastomer sample as a function of time after the laser is turned on, then off for a broad range of temperatures.}\label{fig:TempRaw}
\end{figure}
To estimate the time constant associated with photothermal heating, we measure the back-side temperature of the sample with a thermistor as a function of time after the light is turned on, then blocked.  The thermistor is placed in contact with the sample in its shadow so is not directly exposed to the light, and the same experimental setup used for the photomechanical experiments was used to mount the sample.  The experiment was repeated for an intensity range from about $3.1 \times 10^3 \, \text{W}/ \text{m}^2$ to $2.7 \times 10^4 \, \text{W}/ \text{m}^2$ as shown in Figure \ref{fig:TempRaw}.

The temperature data was fit to a bi-exponential for both the onset phase and relaxation.  The smaller time constant, on the order of 8s to 12s, is associated with a process whose amplitude is a factor of about 10 higher than the slower one, so we take the faster one as being representative of the dominant heating process.  These time constants are plotted in Figure \ref{fig:TimtConstant2and3} as red crosses.  As such, we identify the points on the red fit curve as being associated with the photo-isomerization mechanism and the other points near the red crosses as being associated with photothermal heating.

\begin{figure}\centering
    \includegraphics{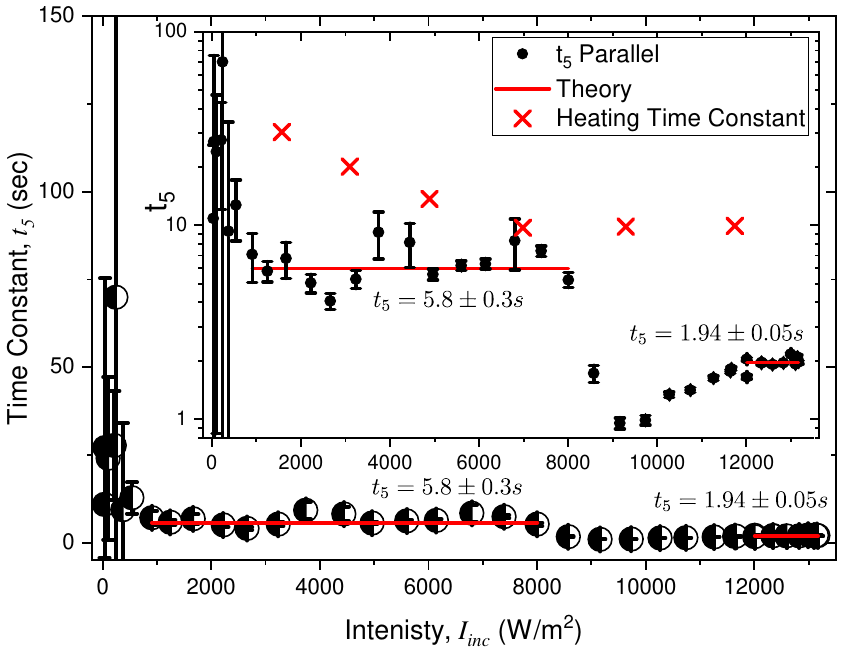}
    \caption{The time constant determined form a fit of the time-dependent stress in a thick liquid crystal elastomer doped with DO1 dye and the theory for relaxation. The inset shows a log plot to magnify the low-amplitude data and the red crosses show the heating time constants determined from the relaxation part of Figure \ref{fig:TempRaw}.}\label{fig:TimeConstant5}
\end{figure}
Figure \ref{fig:TimeConstant5} shows the time constant data for stress relaxation when the beam is turned off as a function of the pump intensity before the light was turned off.  The parameter $t_5$ is associated with the time constant $t_2$ in the fitting function by both sharing the same amplitude $A_2$.  The theory predicts that the relaxation time constant of the photoisomerization mechanism should be given by $t_5 = a$, the parameter $a$ being the same one from Equation \ref{eq:timeConstForm}.  The horizontal red line below intensity $8,000\, \text{W} \cdot \text{m}^{-2}$ is plotted using the parameter $a$ determined from Figure \ref{fig:TimtConstant2and3}.  The inset is a log plot of the same data, and shows that the data agrees well with the predicted time constant.  Furthermore, cooling is associated with a slower process as shown by the red crosses.  Thus, we conclude that the time constant $t_5$ is associated with the direct photo-isomerization mechanism.

Beyond $8,000\, \text{W} \cdot \text{m}^{-2}$  the time constant transitions to another one, suggesting that at higher intensities, another process might be dominating.  However, we place little credence in the higher-intensity data where the material is near the onset of light-induced degradation.  Furthermore, since the sample is thick, heating and photo-isomerization might couple, leading to nonlinear behavior that prevents the response from being separated into a sum of exponentials.  As such, we will not interpret the higher-intensity data.

\begin{figure}\centering
    \includegraphics{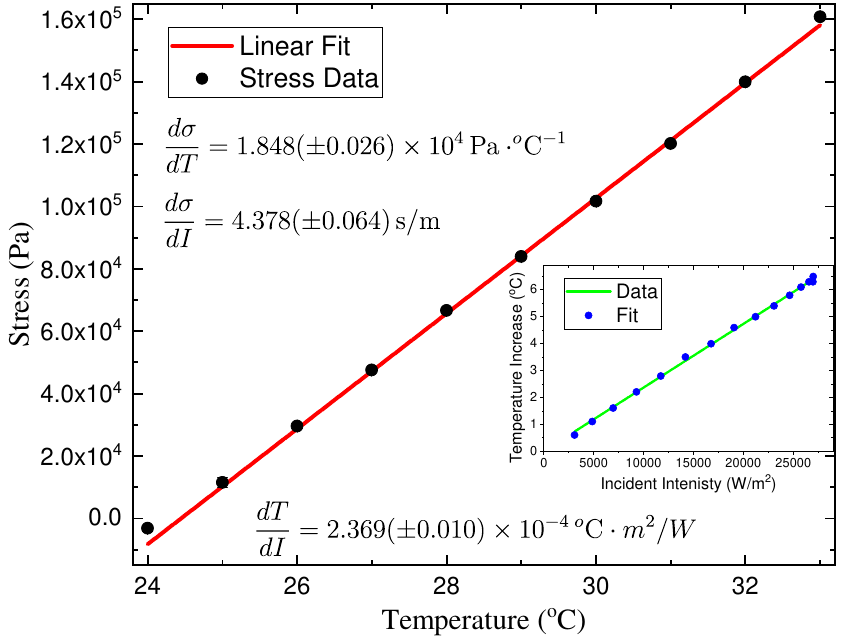}
    \caption{The stress amplitude as a function of temperature and the temperature change as a function of intensity (inset) are used to determine the intensity-dependence of the thermal stress response $d \sigma / d I = 4.378(\pm 0.064) \text{s/m}$.}\label{fig:TempStress}
\end{figure}

The photothermal contribution to the photomechanical stress response can be estimated form the stress induced in a clamped sample in an oven during heating.  The slope of a plot of the stress versus the temperature, as shown in Figure \ref{fig:TempStress}, gives the thermal stress coefficient $d \sigma / dT$.  The inset shows a measurement of the temperature increase of the dark side of the sample after reaching steady state as a function of the incident intensity from the plots shown in Figure \ref{fig:TempRaw}.  Since the sample is thick, the illuminated part of the sample is much hotter than the back side.  The temperature change of the illuminated side is estimated to be over 10 times hotter than the dark side using an infrared camera.

The photomechanical stress response can be calculated from these plots using
\begin{align}\label{eq:calcThermal}
\frac {d \sigma } {d I} = \frac {d T } {d I} \cdot \frac {d \sigma } {d T} = 4.378(\pm 0.064) \, \text{s/m} ,
\end{align}
where we have used the temperature measured in the dark back part of the sample.  If the temperature change of the sample is ten times higher than measured at the back, the photomechanical stress coefficient will be a factor of ten times smaller.

\begin{figure}\centering
    \includegraphics{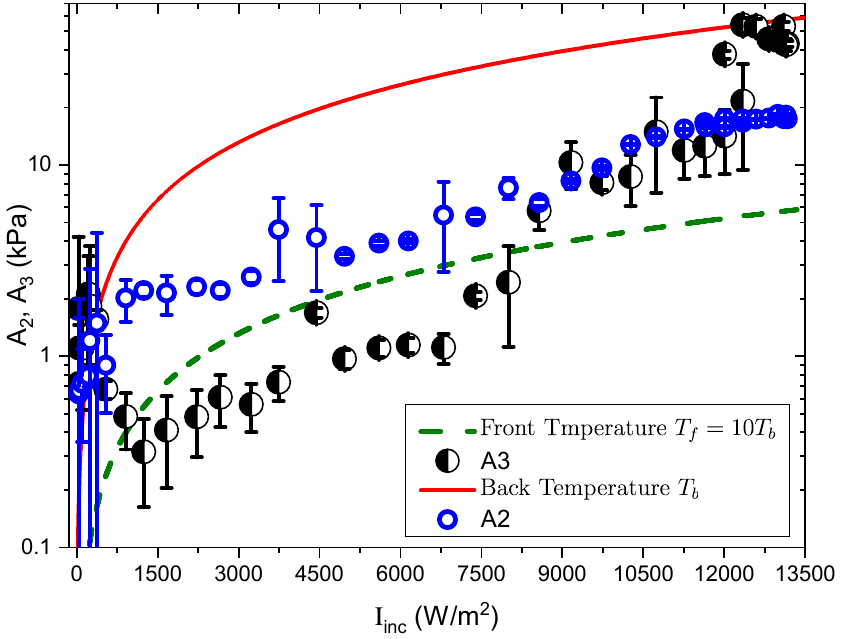}
    \caption{The photoinduced stress amplitudes $A_2$ and $A_3$ as a function of the incident intensity (points) and the theory (curves) using the thermal stress coefficient $\kappa_\sigma^{(1)} = d \sigma / d I = 4.378(\pm 0.064) \text{s/m} $ (solid curve) and $\kappa_\sigma^{(1)}/10 = 0.4378(\pm 0.0064) \text{s/m} $ (dashed curve) determined from Figure \ref{fig:TempStress}.  }\label{fig:PhotoHeatingStress}
\end{figure}
Figure \ref{fig:PhotoHeatingStress} shows a plot of the steady-state stress amplitudes $A_2$ and $A_3$ as a function incident intensity.  The solid theory curve is an overestimate of the photothermal stress response because the sample temperature change is underestimated by about a factor of ten.  The dashed green curve is the theory using ten times the back-side temperature change as an estimate of the average sample temperature.  While the $A_3$ amplitude is somewhat consistent with the dashed theory curve for photothermal heating, this plot alone would not be conclusive proof that $A_3$ represents the photothermal mechanism.  However, it does not rule out our hypothesis that the photothermal response is associated with $A_3$ and that photo-isomerization is connected with $A_2$, as the rest of the data suggests.

\begin{figure}\centering
    \includegraphics{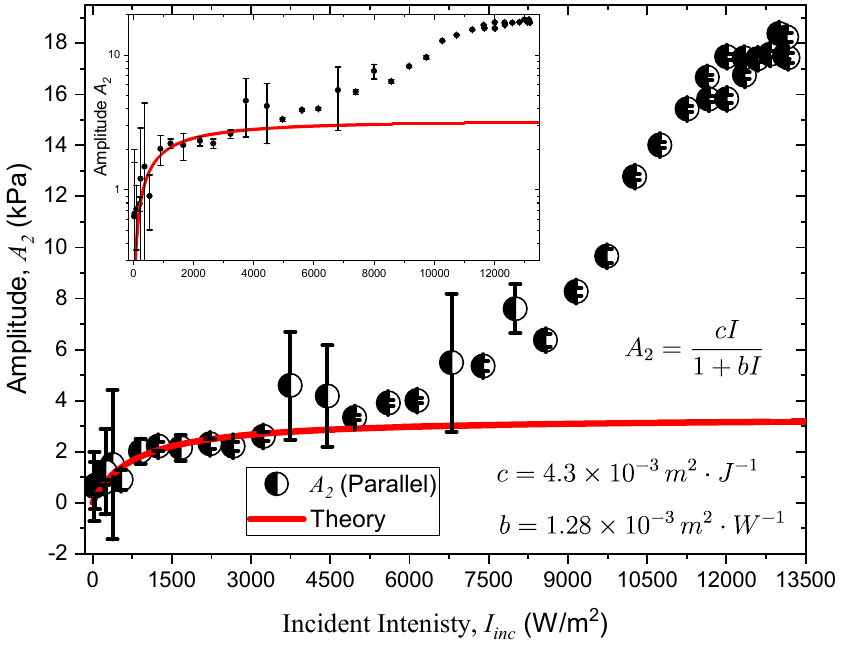}
    \caption{The stress amplitude determined form a fit of the time-dependent stress in a thick liquid crystal elastomer doped with DO1 dye and the theory of an onset process. The inset shows a log plot to magnify the low-amplitude data.}\label{fig:Amplitude2}
\end{figure}
Figure \ref{fig:Amplitude2} shows a plot of the stress amplitude $A_2$ as a function of intensity and the solid red line is a fit to the data to the function
\begin{align}\label{eq:AmplitudeForm}
A_2 = \frac {cI} {1+bI} .
\end{align}
The parameter $b$ determined from the time constant data via Equation \ref{eq:timeConstForm} is fixed while the parameter $c$ is adjusted to fit the data to Equation \ref{eq:AmplitudeForm} at the intensity where the amplitude $A_2$ levels off.  The inset shows a log plot to amplify the lower intensities.  The stress amplitude data also shows that another process becomes dominant above about $6,000\, \text{W} \cdot \text{m}^{-2}$, consistent with the other data.  The plots in this section include the parameters $a$, $b$, and $c$, which will later be used to determine the fundamental parameters of the theoretical models.

We use the polarization dependence of the photomechanical response to separate the heating contribution from photo-isomerization, as first described by Harvey and Terentjev.\cite{harve07.01}  In a polydomain sample, the material is made of microscopic domains of aligned mesogens; but, the domains are aligned randomly with respect to each other.  If the sample is heated and the population of cis isomers grows, the domains will shrink as will the sample.  Under exposure to polarized light, the domains aligned with the light will shrink but those with perpendicular alignment will be unaffected.  As a result, the material will tend to shrink in the direction of the polarized light from photo-isomerization.  To summarize, heating triggers isotropic shrinkage while exposure to polarized light induces an anisotropic response.

\begin{figure}\centering
    \includegraphics{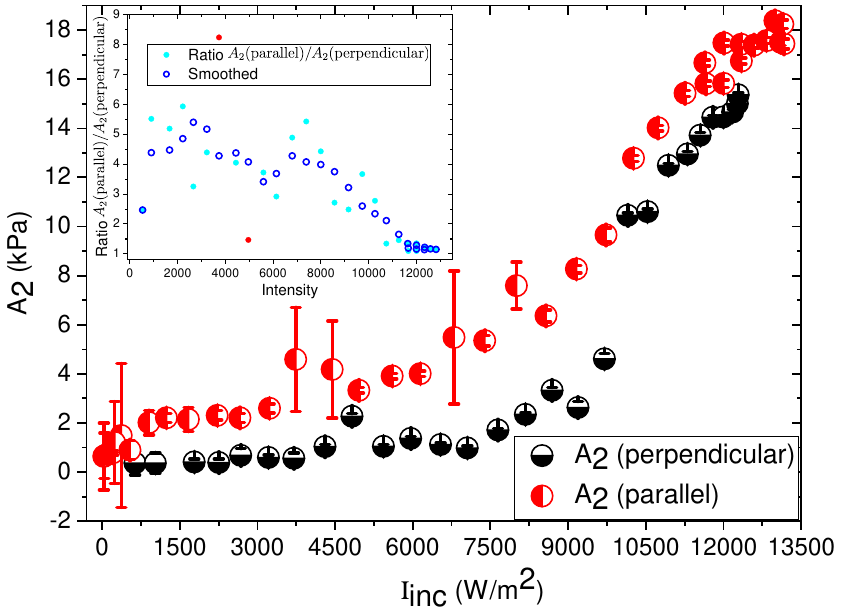}
    \caption{The stress amplitude determined form a fit of the time-dependent stress in a thick liquid crystal elastomer doped with DO1 dye for light polarized parallel to the sample's long axis and perpendicular to it.  The inset shows a ratio of the two.}\label{fig:A2parper}
\end{figure}
Figure \ref{fig:A2parper} shows the dependence of the stress amplitude on the intensity for two polarizations.  A difference between the two responses is an indication of the photo-isomerization mechanism.  The inset shows the ratio between the two.  For low intensities, the signal is highly polarization dependent, suggesting that photo-isomerization is responsible.  At higher intensities, the response becomes polarization independent, suggesting that heating or coupled mechanisms dominate.

Since the parameters $a$, $b$, and $c$ are determined from the low-intensity data, and the observed anisotropy is consistent with photo-isomerization, we will apply the photo-isomerization model in this regime to determine the microscopic photomechanical parameters.  Note that the thickness of the sample and the noise in the signal makes an error analysis meaningless and the following microscopic values determined from data should be taken as an order-of-magnitude estimate.

A comparison between Equation \ref{eq:PopTimeConstInfinity} and Equation \ref{eq:timeConstForm} immediately gives
\begin{align}\label{eq:beta}
\beta = 1/a = 0.17\, \text{s}^{-1}
\end{align}
and
\begin{align}\label{eq:alpha}
\alpha = b/a = 2.2 \times 10^{-4} \, \text{m}^2 \cdot \text{J}^{-1} .
\end{align}

To get the microscopic parameters form the data, we make two assumptions that greatly simplifies the analysis.  First, we assume that the spring constant of the excited spring and the resting spring are the same, or $k^\prime = k$.  Furthermore, we assume that light-induced de-excitation rate of the spring is negligible compared with photo-induced excitations.  Then, $\alpha=\alpha_0$,  and Equation \ref{eq:StressResponseBeforeExpand} yields
\begin{align}\label{eq:StressResponseApprox}
\sigma(t \rightarrow \infty) &= \frac {\frac {\alpha} {\beta} \frac { \left( x_b - a \right)} {\ell} E  I } {1 - \frac {\alpha} {\beta} I} .
\end{align}
A comparison of Equation \ref{eq:AmplitudeForm} and Equation \ref{eq:StressResponseApprox} yields
\begin{align}\label{eq:StressResponseC}
c &=  \frac {\alpha} {\beta} \frac { \left( x_b - a \right)} {\ell} E ,
\end{align}
so with $E=1.26 \, \text{MPa}$ and $\alpha$ and $\beta$ given by Equations \ref{eq:alpha} and \ref{eq:beta} with $c=4.3 \times 10^{-3} \, \text{m}^2 \cdot \text{J}^{-1}$ yields
\begin{align}\label{eq:PMStrain}
 \frac { \left( x_b - a \right)} {\ell} = 2.6 \times 10^{-6} .
\end{align}

The DO1 dye fraction in the elastomer is estimated to be $10^{-3}$ as determined from absorption spectroscopy.\cite{camac04.01}  Since the DO1 molecule's mass is about four times the mass of a polymer unit, there are $4\times 10^3$ polymer links per dye molecule.  Given the length of a link of about $3$\AA,\cite{Mark09.01} the average polymer chain length per dye molecule is $\ell = 1.2 \times 10^4$\AA, leading from Equation \ref{eq:PMStrain} to a molecular length change of
\begin{align}\label{eq:MoleculeLengthChange}
 \left( x_b - a \right) = 0.03 \, \text{\AA}.
\end{align}

The length difference value of $0.03$\AA\hspace{0.1em} is about a thirtieth of the approximate $1$\AA \hspace{0.1em} length difference between isomers, suggesting a 3\% efficiency of the length change being transferred to a force acting on the material.  There are many possible sources of the observed efficiency: (1) According to Equation \ref{eq:PMunitEffic}, the azo-dye is hindered by its environment, and needs to overcome the inertia of the mesogens in the domain where it is found; (2) the strength of liquid crystal interactions;\cite{tosch14.01}  (3) not all molecules are coupled to domains; and (4) the photo-isomerization quantum efficiency is not 100\% upon light absorption.  Thus, the observed microscopic photomechanical efficiency is within the expected range, and considering the level of approximation and the order-of-magnitude nature of the measurements, a reasonable result.

\begin{table}
\begin{tabular}{c c c}
  \hline
   Parameter & Description & Value \\ \hline
 $\alpha$ & Conversion Efficiency  & $2.2 \times 10^{-4} \, \text{m}^2 \cdot \text{J}^{-1} $  \\
 $\beta$ & Relaxation Rate & $0.17\, \text{s}^{-1} $ \\
 $\ell$ & Photomorphon Length & $1.2 \times 10^{-6} \, \text{m}$ \\
 $\left( x_b - a \right) / \ell $ & Photomorphon Strain & $2.6 \times 10^{-6}$ \\
  $x_b - a $ & Dye Length Change & $0.03 \times 10^{-10} \, \text{m}$ \\
  \hline
\end{tabular}\caption{A summary of the microscopic parameters determined from the measurements of DO1-doped PDMF liquid crystal elastomer..\label{tab:paramters}}
\end{table}
Table \ref{tab:paramters} summarizes the order-of-magnitude microscopic parameters determined from the set of measurements.  These values for various materials can be used to design new materials that better leverage these properties.  An important future direction of research is accurate characterization of materials for both materials comparisons, model validation and guiding future material and device development.

\subsubsection{Figure of Merit}

The last column of Table \ref{tab:results} summarizes the figures of merit for the materials measured.  The monodomain liquid crystal elastomer's FOM is comparable to that of the dye-doped PMMA fibers and handily beats the polydomain material.  It is interesting to note that the stress response of the PMMA fibers is by far the largest, but the larger Young's modulus of the bulk lowers the FOM.  The stress response of the monodomain LCE is small, but its low Young's modulus makes the figure of merit large.

It is intriguing that the FOM is comparable for both elastomer and glassy polymer hosts given that the mechanisms of the response are different.  As discussed in Section \ref{sec:overview}, the heating mechanism's efficiency is low, so it appears that the photo-isomerization mechanism as leveraged by liquid crystalline domains provides little advantage.  This suggests that there is much room for improvement in the design of new materials, but novel paradigms will be required.

\subsection{Hierarchy of Efficiencies}

There is a hierarchy of inefficiencies of the photomechanical response, each process dissipating energy that does not lead to work.  As described in the introduction, a fundamental limit to the efficiency starts at the molecular level as described by how much of the photon's absorbed energy is converted to work.  However, not all absorbed photons change the molecule's shape and thus do not contribute to work.  While the absorbed energy will often be dissipated as heat, heating leads to thermal expansion so it still adds to the work done.  Thermal expansion is in principle inefficient, though in many materials, it is the dominant mechanism.  For the purposes of understanding the potential for harnessing molecular forces, we neglect photothermal heating in determining molecular efficiency.  Putting it all together, the molecular efficiency is given by
\begin{align}\label{eq:MolEff}
Q_e^\text{mol} = \xi \left(1 - \frac {V_0 } {\hbar \omega} \right) ,
\end{align}
where $V_0$ is the energy dissipated as the system relaxes to its initial state and $\xi$ is the fraction of absorbed photons that do work.

Equation \ref{eq:PMunitEffic} accounts for a further limit on the efficiency due to interactions of the active molecule with the surrounding material.  Then, the efficiency at the level of the photomorphon is given by
\begin{align}\label{eq:PMunitEfficTot}
Q_e^\text{PM} = \xi_\text{PM} Q_e^\text{mol} =  \xi \frac {k_b} {k_e + k_b}  \left(1 - \frac {V_0 } {\hbar \omega} \right) .
\end{align}
If a material is assembled in a way that no parasitic spring forces act in parallel with the active molecule and a molecule is designed to minimize $V_0$, then the efficiency at the level of the photomorphon is limited solely by the Franck-Condon factor.  This interplay between the various factors might be controllable through quantum design, a topic that deserves attention.  The salient point is that optimization of these factors is possible, and that huge enhancements over the best materials at present might result with clever manipulations on the microscopic scale.

Next we consider the work done at the macro-scale of the loop shown in Figure \ref{fig:IxCycle}.  Recall that we evaluated the loop in a counterclockwise path because we were using an external force to characterize the photomorphon.  To convert light into work, we need to run the loop in the clockwise direction.

Starting on the left segment, the light imparts energy to the spring in the amount
\begin{align}\label{eq:dU4}
\delta U_4 =  \frac {1} {2} k^\prime \left( x_0 - x_0^\prime \right)^2
\end{align}
as the spring constant and equilibrium length changes.  Since there is no displacement, no work is done.  On the top segment, the work on the environment by the spring is given by
\begin{align}\label{eq:dW3}
\delta W_3 = \frac {1} {2} k^\prime \left( x_0 - x_0^\prime \right)^2 - \frac {1} {2} k^\prime \left( x - x_0^\prime \right)^2  ,
\end{align}
where an external agent brings the spring to length $x$.  If the spring relaxes to its equilibrium length, $x = x_0^\prime$.  On the righthand segment, no work is done but energy is released from the spring to light or heat in the amount
\begin{align}\label{eq:dU2}
\delta U_2 = \frac {1} {2} k \left( x - x_0 \right)^2 - \frac {1} {2} k^\prime \left( x - x_0^\prime \right)^2 .
\end{align}
Finally, the spring relaxes to its equilibrium resting length while doing work
\begin{align}\label{eq:dW1}
\delta W_1 = \frac {1} {2} k \left( x - x_0 \right)^2.
\end{align}

The total work done by the spring on the environment is give by
\begin{align}\label{eq:TotalW}
\delta W = \delta W_1 + \delta W_3
\end{align}
and the net energy supplied by the light is
\begin{align}\label{eq:TotalU}
\delta U = \delta U_2 + \delta U_4 .
\end{align}
It is easy to verify that energy is conserved, or
\begin{align}\label{eq:Econcserve}
\delta U = \delta W.
\end{align}

\subsubsection{Single Photon Processes}

For illustration, we evaluate the loop for the special case of two single-photon absorptions as shown in Figure \ref{fig:SpringsModel} for a molecule.  This corresponds to the path given by the vertical arrow labelled $\hbar \omega$, then relaxation along the upper parabola, the excitation by the vertical arrow labelled $\hbar \omega^\prime$ followed by a relaxation along the left parabola to the resting state, as follows.

Excitation with the the first photon changes the spring constant to $k^\prime$ with the length remaining at $x = x_0$ and the Equilibrium length changing to $x_0^\prime$, yielding from Equation \ref{eq:dU4}
\begin{align}\label{eq:dU4phpton}
\delta U_4 =  \frac {1} {2} k^\prime \left( x_0 - x_0^\prime \right)^2 = \hbar \omega .
\end{align}
Next, Equation \ref{eq:dW3} describes the part of the cycle that does work, yielding
\begin{align}\label{eq:dW3photon}
\delta W_3 = \frac {1} {2} k^\prime \left( x_0 - x_0^\prime \right)^2 = \hbar \omega - V_0
\end{align}
when the system reaches the excited-state length $x = x_0^\prime$.  Upon absorbing the next photon of energy $\hbar \omega^\prime$ to excite the system onto the ground-state spring potential energy surface, the energy change given by Equation \ref{eq:dU2} is
\begin{align}\label{eq:dU2photon}
\delta U_2 = \frac {1} {2} k \left( x_0^\prime - x_0 \right)^2 = \hbar \omega^\prime,
\end{align}
where the length remains fixed at $x = x_0^\prime$ during excitation.  Finally, according to Equation \ref{eq:dW1}, the system does work as it comes back to its equilibrium resting length
\begin{align}\label{eq:dW1photon}
\delta W_1 = \frac {1} {2} k \left( x_0^\prime - x_0 \right)^2 = \hbar \omega^\prime + V_0.
\end{align}

In this process, the work done by the system on the environment is the absorbed photon energy as given by Equation \ref{eq:TPAWorkDone}, as can be verified by evaluating Equations \ref{eq:TotalW} and \ref{eq:TotalU}, which we can verify by adding Equations \ref{eq:dW3photon} and \ref{eq:dW1photon}, yielding
\begin{align}\label{eq:dWTotphoton}
\delta W = \frac {1} {2} \left(k + k^\prime \right) \left( x_0 - x_0^\prime \right)^2 = \hbar \omega + \hbar \omega^\prime .
\end{align}
Note that the photon energies are not arbitrary but are chosen to specifically match the vertical transitions.  As such, Equation \ref{eq:dWTotphoton} relates those molecular energies to the spring constants and their equilibrium lengths.  Subtracting Equation \ref{eq:dU2photon} from \ref{eq:dW1photon} shows that we have implicitly chosen the most efficient molecular process with $V_0=0$.  As such, the efficiency is limited only by the branching ratio of excitations that result in a change of length and spring constant compared to other processes.

Finally, we consider the one-photon process.  The first two steps are the same as that given by Equations \ref{eq:dU4phpton} and \ref{eq:dW3photon}.  In the final tunneling process, heat $\delta Q$ is dissipated in the last step with
\begin{align}\label{eq:HeatDiss}
\delta Q = V_0 .
\end{align}
The efficiency of the cycle, if powered by light at frequency $\omega$ is given by Equation \ref{eq:PMunitEfficTot}.

\subsubsection{Equilibrium Processes}

The excited state remains populated only when the material is illuminated.  The higher the intensity, the greater the excited state equilibrium population fraction, as give by Equation \ref{eq:PopTimeInfinity}.  Unlike the single photon/molecule interaction, where the photon energy is converted to work done by the molecule, the light must continuously supply power to keep the system in the desired state of equilibrium.  As such the efficiency increases when the work cycle is traversed with minimum possible time, with the single-photon process as the limiting case of ultimate efficiency.

Whereas the single-molecule picture focuses on one photon interacting with one atom, the statistical model depends on populations that are modelled by the average photomorphon.  Then, the ground state of the ensemble consists of a photomorphon in its ground state and the excited state is that of the average photomorphon, which is in a combination of states, weighted by the populations of the ground and excited molecules.

We start by evaluating Equation \ref{eq:PopTimeTurnOn} in the short time limit ($\beta + \alpha I) t \ll 1$) so that the exponential can be expanded in a series.  Furthermore, we assume that the temperature is low enough compared with the energy difference between the two states of the springs such that the excited state population vanishes with no light, or $n_e^{(0)} = 0$.  Then, Equation \ref{eq:PopTimeTurnOn} becomes
\begin{align}\label{eq:PopTimeTurnOnSmallt}
n_e (t) = \frac { \alpha_0 I -  \alpha _0 I \left( 1 - (\beta + \alpha I)t \right)} {\beta + \alpha I} = \alpha I t.
\end{align}

Equation \ref{eq:PopTimeTurnOnSmallt} can then be used to determine the spring constant of the average photomorphon using Equation \ref{eq:AverageChainSpringConstant}
\begin{align}\label{eq:AverageChainSpringConstantSmallt}
K(t,I) =  \frac {k} {1 - \alpha I t \left( 1 - \frac {k} {k^\prime} \right) } \approx  k \left(1 - \alpha I t \left( 1 - \frac {k} {k^\prime} \right) \right).
\end{align}
where we recover the result $K=k$ at $t=0$, at the instant the light is turned on.  Similarly, the displacement will be given by Equation \ref{eq:PopDependStrain}
\begin{align}\label{eq:PopDependStrainSmallt}
\delta \ell(t,I) = \alpha I t (x_0^\prime - x_0 ) ,
\end{align}
with the force being derived from Equations \ref{eq:AverageChainSpringConstantSmallt} and \ref{eq:PopDependStrainSmallt}, yielding to first-order in time the expression
\begin{align}\label{eq:ForceSmallt}
F(t,I) = K(t,I) \cdot \delta \ell(t,I) = k \alpha I t (x_0^\prime - x_0 ) .
\end{align}
Thus, to lowest order in time, the force is solely due to the change in the spring length.

To evaluate the loop given by Figure \ref{fig:IxCycle},  $K(t,I)$ is the excited photomorphon so is associated with $k^\prime$ in Equations \ref{eq:dU4}-\ref{eq:dW1} and $\delta \ell$ is associated with $x_0^\prime - x_0$.  Thus, setting $x = x_0 + \delta \ell$, Equation \ref{eq:dW3} to lowest order in time becomes
\begin{align}\label{eq:w3pop}
\delta W_3 = \frac {1} {2} k \alpha^2 I^2 t^2 \left( x_0^\prime - x_0 \right)^2
\end{align}
and Equation \ref{eq:dW1} yields $\delta W_1 = \delta W_3$.
Note that Equation \ref{eq:w3pop} seems to imply that the amount of work provided by the light grows indefinitely.  This is an artifact of using the short-time approximation, at which point the molecular populations are evolving.  After the excited state population comes to equilibrium, no additional work will be done, and the energy from the light beam will be wasted.

Using the time constant given by Equation \ref{eq:PopTimeConstInfinity} for the effective time over which work is done on the environment, Equation \ref{eq:w3pop} with $\delta W_1 = \delta W_3$ gives
\begin{align}\label{eq:WorkDone}
\delta W  =  k \alpha^2 I^2 \left( \frac {1} {\beta + \alpha I} \right)^2 \left( x_0^\prime - x_0 \right)^2 .
\end{align}
Over this time, the energy absorbed by a sample with a face of area $A$ that is fully illuminated is $I A \tau$, so the efficiency is then given by
\begin{align}\label{eq:WorkEff}
\xi_I = \frac {\delta W} {IA \tau} =  \frac {k \alpha} {A} \left( \frac {\alpha I} {\beta + \alpha I} \right) \left( x_0^\prime - x_0 \right)^2 .
\end{align}
Expressing Equation \ref{eq:WorkEff} in terms of Young's modulus $E$ yields
\begin{align}\label{eq:WorkEffBulk}
\xi_I =  \frac {E  \alpha} {\ell} \left( \frac {\alpha I} {\beta + \alpha I} \right) \left( x_0^\prime - x_0 \right)^2 .
\end{align}

In the low-intensity regime, where $\alpha I \ll \beta$, the term in parenthesis in Equation \ref{eq:WorkEffBulk} is also much smaller than unity.  In the high-intensity limit,  where $\alpha I \gg \beta$, the term in parentheses is unity.  Thus, the largest photomechanical efficiency is at high intensity and is given by
\begin{align}\label{eq:WorkEffBulkMax}
\xi_I =  \frac {E  \alpha} {\ell} \left( x_0^\prime - x_0 \right)^2 .
\end{align}

Equation \ref{eq:WorkEffBulkMax} is the high-intensity limit of optimum exposure time that takes advantage of the macroscopic population change, but is long enough to avoid quantum fluctuations associated with single photon processes.  This is the realm of high-efficiency practical applications.

We can evaluate this efficiency for the DO1-doped liquid crystal elastomer using the experimentally-determined microscopic parameters in Table \ref{tab:paramters} and Young's modulus from Table \ref{tab:results} noting that $ x_0^\prime - x_0 = x_b -a$ is the average change of the photomorphon length.  This yields an efficiency of $\xi_I = 2.1 \times 10^{-15}$.  This result shows that polydomain liquid crystal elastomers are highly inefficient.  The monodomain liquid crystal elastomer's Young's modulus is double the value of the polydomain material, and the photomechanical FOM is about 10 times great, leading to an efficiency of perhaps two orders of magnitude, still 13 orders of magnitude below what is possible.  Dye-doped PMMA might add another three-orders of magnitude in efficiency, still 10 orders of magnitude short.

This analysis shows that the photomechanical efficiency of present-day materials is far from ideal, so new materials paradigms will be required to efficiently harness the mechanical energy of light.  The measurement techniques and analysis presented in this paper can be used to guide the development of new materials by suggesting how the bulk and microscopic parameters need to be tuned and providing a measurement technique that characterizes the critical parameters.

\section{Conclusions}

Our work melds time-dependent measurements of the photomechanical stress response with phenomenological models of the underlying mechanisms to provide a unified view for teasing out the origins of light-induced stress and determining its efficiency.  The microscopic photomechanical response originates in the photomorphon, which is the fundamental engine of the response.  We showed how it can be an active molecule that changes shape when absorbing a photon, or a cell made of a polymer chain containing mesogens and a light-activated molecule.  The photomorphon is modeled based on the structure of the material, and its response dictates the observed bulk stress or strain response.  Conversely, we showed how bulk measurements can be used to determine the microscopic parameters, which can be used to test the models and guide the development of new materials.

We model a molecule as a spring whose force constant and resting length changes upon excitation.  A passive spring in parallel with the active molecule forms the photomechanical unit, and when embedded in series with a polymer forms the photomorphon.

Photothermal heating and photo-isomerization of dopant molecules are shown to exhibit the same universal behavior.  Auxiliary measurements, such as light-induced temperature change and temperature-dependent stress, must be used to decouple bulk thermal contributions, attributing the remainder to other mechanisms.  We applied this approach to isolate the photoisomerization mechanisms, and showed that the microscopic parameters -- such as the average length change of the molecule -- are in the right ballpark.  The data analysis methods presented show pitfalls in the fitting process and how they can be avoided, and these were illustrated in separating the heating and isomerization response.

The bulk figure of merit was determined on the macro scale for various materials such as liquid-crystal elastomers doped with molecules whose confirmations changes upon photo-excitation, polymers doped with molecules that change shape when excited with light, and inactive dopant molecules that only absorb light that heats the material.  Dye-doped polymers were found to have the largest FOM while being dominated by the thermal response while dye-doped liquid crystal elastomers have lower figures of merit and combine both photothermal heating and photo-isomerization mechanisms.

This paper has focused mostly on polymeric materials. However, molecular crystals offer great promise due to their high density of photomechanical molecules.\cite{naumov14.02,naumov14.01,ISI:000331779800015,naumov15.01,naumov13.01}  Crystals therefore have the potential for a much larger photomechanical response due the absence of parasitic passive components and higher young's modulus.  The theory presented above would apply with the unit cell acting as the photomorphon.

In summary, we have presented an experimental protocol for determining the microscopic parameters that govern the bulk photomechanical response.  These parameters, which can be associated with various components of the material, are visualized as a composite made of springs that defines the photomorphon, whose length and spring constant changes upon light excitation.  A population model relates the microscopic properties of the photomorphon to the dynamics of the bulk material, making it possible to fully characterize the parameters by measuring the stress as a function of time, temperature and intensity.   A hierarchy of figures of merit is presented, and the best FOM for photomechanical efficiency is found to fall many orders of magnitude short to what is possible.  The experiments, theory and method of analysis presented here can be used to study the mechanisms of the response,  characterize materials and to develop new ones with properties that are optimized for a given application.

{\bf Acknowledgements} We thank the National Science Foundation (EFRI-ODISSEI:1332271) for supporting this work; and, Peter Palffy-Muhoray and Misha Pevnyi for supplying elastomer and carbon nanotube samples.

\end{document}